\documentclass{emulateapj}

\usepackage{amsmath}
\usepackage{amssymb}
\usepackage{amsfonts}
\usepackage{amstext}
\usepackage{graphicx}
\usepackage{fancybox}
\usepackage[usenames,dvipsnames]{color}
\usepackage{pstricks}
\usepackage{bm}
\usepackage{soul, color}
\usepackage{latexsym}
\usepackage{pifont}
\usepackage{shorttoc}
\usepackage{subfigure}
\usepackage{textcomp} 
\usepackage{float}
\usepackage{enumitem}

\usepackage{bm}
\usepackage{lipsum}
\usepackage{threeparttable}
\usepackage{hyperref}
\hypersetup{urlcolor=cyan}

\def\aap{A\&A}

\def\apjs{ApJS}
\def\apjl{ApJL}

\def\lsim{~\rlap{$<$}{\lower 1.0ex\hbox{$\sim$}}}

\def\gsim{~\rlap{$>$}{\lower 1.0ex\hbox{$\sim$}}}
\newcommand{\myemail}{l.m.ribas@astro.uio.no}

\shorttitle{Boosting Ly$\alpha$ and H\lowercase{e}II$\,\lambda$1640 in 
stochastic Pop III galaxies}
\shortauthors{Ll. Mas-Ribas et al.}

\begin{document}

\title{Boosting Ly$\alpha$ and H\lowercase{e}II$\,\lambda$1640 line fluxes from Population III 
galaxies:\\  stochastic IMF sampling and departures from case-B}

\author{Llu\'is Mas-Ribas\altaffilmark{1}} 
\author{Mark Dijkstra\altaffilmark{1}}
\author{Jaime E. Forero-Romero\altaffilmark{2}}
\altaffiltext{1}{Institute of Theoretical Astrophysics, University of Oslo,
Postboks 1029, 0315 Oslo, Norway ~  \url{\myemail} }
\altaffiltext{2}{Departamento de F\' isica, Universidad de los Andes, Cra. 
1 No. 18A-10, Edificio sIp, Bogot\'a, Colombia}

\begin{abstract}

We revisit calculations of nebular hydrogen Ly$\alpha$ and ${\rm HeII}\,\lambda1640$ line 
strengths for population III galaxies, undergoing continuous and bursts of star formation. 
We focus on initial mass 
functions (IMFs) motivated by recent theoretical studies, which generally span a lower 
range of stellar masses than earlier works. We also account for case-B departures and the stochastic 
sampling of the IMF.
In agreement with previous works, we find that departures from case-B can 
enhance the Ly$\alpha$ flux by a factor of a few, but we argue that this enhancement is driven 
mainly by collisional excitation and ionization, and not due to photoionization from the $n=2$ 
state of atomic hydrogen. The increased sensitivity of the Ly$\alpha$ flux to the high-energy 
end of the galaxy 
spectrum makes it more subject to stochastic sampling of the IMF. The latter introduces a dispersion 
in the predicted nebular line fluxes around the deterministic value by as much as a factor of $\sim 4$.  
In contrast, the stochastic sampling of the IMF has less impact on the emerging Lyman Werner 
(LW) photon flux. 
When case-B departures and stochasticity effects are combined, nebular line emission 
from population III galaxies can be up to one order of magnitude brighter than predicted 
by `standard' calculations that do not include these effects. This enhances the 
prospects for detection with future facilities such as JWST and large, groundbased telescopes.
\end{abstract}


\section{Introduction}

  The first generation of stars, so called Population III (Pop III; hereafter) stars, played an important 
role in setting up the `initial conditions' for galaxy formation in our Universe: Pop III stars initiated 
Cosmic Reionization and deposited the first heavy elements into the interstellar (ISM) and 
intergalactic media (IGM). In addition, Pop III stars likely provided seeds for the formation of the 
massive black holes we observe today \citep[see, e.g., the reviews by][]{Bromm2013,Karlsson2013,Greif2015}. All these processes, in detail, depend on the physical 
properties of Pop III stars, such as their mass, temperature, spin, and stellar evolution.

  Most predictions of observational signatures of Pop III stars have focused on their spectral 
energy distribution (SED) \citep[see][for a detailed review]{Schaerer2014}. One of the most 
important parameters affecting the SED is the stellar mass, since it strongly affects the stars' total 
and ionizing luminosities. Unfortunately, the mass of Pop III stars, as well as the 
initial mass function (IMF), is still uncertain. Early works 
focused on the study of very massive objects \citep[several hundreds of solar masses; e.g.,][]
{Bromm2001,Bromm2002,Abel2002,Schneider2002,Schneider2003}. Although it is still possible 
to form stars with masses of a few hundreds of solar masses  in current calculations
\citep[e.g.,][see also \cite{Krumholz2009}]{Omukai2002,Mckee2008, Kuiper2011,
Hirano2014}, it is also likely to obtain low mass stars, sometimes of only a few tens 
of solar masses or less \citep[e.g.,][]{Turk2009, Stacy2010,
Greif2011,Clark2011,Stacy2012,Stacy2014,Susa2014,Hirano2015,Hosokawa2015}. 
The preference for masses less than $\sim150$ M$_{\odot}$ is  
consistent with abundance patterns found in extremely 
metal-poor stars  \citep[e.g.,][see also \cite{Keller2014}]{Umeda2005,Chen2016}, metal-poor 
Damped Lyman Alpha systems\footnote{Damped Lyman Alpha systems (DLAs) are absorption 
systems with hydrogen column densities above $2\times10^{20}\,{\rm cm^{-2}}$, where the gas in 
the core is in the atomic form due to self-shielding from the ionizing background radiation 
\citep[see][for a detailed review]{Wolfe2005}.} \citep{Pettini2002,Erni2006} and with the inferred  
number of pair instability supernovae \citep[PISN;][]{Heger2002} events \citep[e.g.,][]{Tumlinson2004, 
Tumlinson2006,Karlsson2008,Frebel2009,Aoki2014, Bennassuti2014,Frebel2015}.

  Direct searches for Pop III stars/galaxies have mostly focused on detecting strong 
hydrogen Ly$\alpha$ and ${\rm HeII}\, \lambda1640$ line emission 
\citep[e.g.,][]{Schaerer2008,Nagao2008}, the latter being associated with the spectrum of {\it 
massive} Pop III stars \citep[e.g.,][]{Tumlinson2000,Tumlinson2001,Bromm2001,Malhotra2002,Jimenez2006,
Johnson2009,Raiter2010}. Interestingly, \citet{Sobral2015} have recently reported on the discovery of 
a high-redshift galaxy with unusually large Ly$\alpha$ and ${\rm HeII}\, \lambda1640$ line 
emission, which had led to speculation that we might have discovered a massive Pop III galaxy, 
although this is still a matter of intense 
debate \citep{Pallottini2015,Dijkstra2015,Smidt2016,Smith2016,Visbal2016,Xu2016}.

  In this work, we revisit calculations of the spectral signature of Population III galaxies. Our 
work differs from previous analyses in the following ways: (\textit{i}) We 
allow for stellar populations with lower stellar mass limits than previous studies. This is motivated by 
the more recent theoretical and observational preferences for lower mass Pop III stars.  
(\textit{ii}) We examine, for the first time, the effects of the stochastic sampling of the IMF to 
the photon flux and luminosities from Pop III stellar populations. As we will show, stochasticity  
effects can be significant, and they can be further amplified when departures from case-B 
recombination assumption are taken into account. 

Our paper is structured as follows: In Section \ref{sec:models} we show the calculations to 
model the stellar and nebular SEDs, and we also detail the method for the stochastic sampling 
of the IMF. We present and analyse our results in Section \ref{sec:results}, and we discuss them 
in Section \ref{sec:discussion} before concluding in Section
\ref{sec:summary}.

\section{Stellar atmosphere, population and nebula models}\label{sec:models}

  In this section, we present the calculations that yield to
the obtention of the final spectra.  We compute stellar atmosphere SEDs   
in Section \ref{sec:stars}. In Section \ref{sec:pops} we construct several stellar 
populations accounting for various IMFs and the stochastic sampling of the IMF. 
In Section \ref{sec:nebula} 
we include the nebula surrounding the stellar populations using the photoionization 
code \textit{Cloudy} version 13.03 \citep{Ferland2013}.

  Hereafter, we will refer to stellar populations containing low-mass Pop III stars when 
the upper limit of the IMF is  $\leq 100\,{\rm M_{\odot}}$, and high-mass star populations
for upper limits above this threshold. 
\subsection{Stellar atmosphere models}\label{sec:stars}

  We use the code \textit{Tlusty} v200 \citep{Hubeny1995} for the creation of the stellar SEDs. 
This code performs radiative transfer calculations in stellar atmospheres allowing for line 
blanketing and non-LTE effects. The latter is of particular importance in very hot 
stars, where the radiative processes are dominant 
\citep[e.g.,][]{Marigo2001,Schaerer2002a,Bromm2002,Kubat2012,Rydberg2013,Lovekin2014}. 
The code considers a plane-parallel geometry, which is also a good approximation in our case, since 
Pop III stars have large masses and radii \citep[][]{Schaerer2002a,Kubat2012}. 

\begin{table*}
	\begin{center}
	\caption{Stellar model atmosphere parameters\label{ta:starparams}}	
	\begin{threeparttable}
		\begin{tabular}{cccccccccc} 
		\hline
		\hline
$M\,(M_{\odot})$    &$T_{eff}\,{\rm (K)}$    &$R\,(R_{\odot})$   &${\rm log}\,g$    &lifetime (${\rm Myr}$)    &${\rm Q\,(HI)}\tnote{b}  $	   &${\rm Q\,(HeII)}\tnote{b}  $        &${\rm Q\,(LW)}\tnote{b}  $   	&$\overline{{\rm E}}$\tnote{c}	&Source\tnote{a}          \\ 
		\hline  
		        $1\,000$      &$106\,170$    &$15.57$     &$5.053$    &$1.59$    	&${\rm1.54\,x\,10^{51}}$        &${\rm3.51\,x\,10^{50}}$        	&${\rm1.08\,x\,10^{50}}$   &$2.99$   &sch02   \\  
		        $500$      &$106\,900$   &$10.41$     &$5.102$       &$1.89$  	&${\rm7.16\,x\,10^{50}}$        &${\rm1.55\,x\,10^{50}}$        	&${\rm5.12\,x\,10^{49}}$    &$2.95$  &sch02 \\  
		          $400$      &$106\,660$     &$9.09$    &$5.123$       &$1.97$ 		&${\rm5.47\,x\,10^{50}}$        &${\rm1.13\,x\,10^{50}}$        	&${\rm3.98\,x\,10^{49}}$  &$2.91$	 &sch02 \\ 
		         $300$      &$101\,620$     &$8.28$   &$5.079$         &$2.07$  	&${\rm3.90\,x\,10^{50}}$        &${\rm6.75\,x\,10^{49}}$        	&${\rm3.13\,x\,10^{49}}$  &$2.77$	 &sch02     \\  
		         $200$      &$99\,770$      &$6.48$  &$5.116$         	&$2.24$	 &${\rm2.27\,x\,10^{50}}$        &${\rm3.29\,x\,10^{49}}$        	&${\rm1.93\,x\,10^{49}}$   &$2.67$	 &sch02    \\  
		           $120$      &$95\,720$    &$4.81$    &$5.153$        	&$2.55$	 &${\rm1.11\,x\,10^{50}}$        &${\rm1.21\,x\,10^{49}}$        	&${\rm1.03\,x\,10^{49}}$   &$2.53$	 &sch02   \\ 
		        $100$      &$94\,400$      &$4.23$   &$5.185$           &$2.70$	        	&${\rm8.26\,x\,10^{49}}$        &${\rm7.59\,x\,10^{48}}$        	&${\rm7.96\,x\,10^{48}}$  &$2.48$	 &mrg01\\  
		        $80$      &$93\,320$      &$3.60$   &$5.230$         	&$2.93$	 &${\rm5.79\,x\,10^{49}}$        &${\rm4.50\,x\,10^{48}}$        	&${\rm5.74\,x\,10^{48}}$   &$2.44$   &sch02  \\  
		         $70$      &$89\,530$      &$3.44$  &$5.209$           	&$3.10$        	&${\rm4.59\,x\,10^{49}}$        &${\rm2.50\,x\,10^{48}}$        	&${\rm4.94\,x\,10^{48}}$  &$2.35$	&mrg01 \\  
		        $60$      &$87\,700$       &$3.12$  &$5.228$         	&$3.33$	&${\rm3.53\,x\,10^{49}}$        &${\rm1.42\,x\,10^{48}}$        	&${\rm3.97\,x\,10^{48}}$    &$2.30$   &sch02  \\  
		         $50$      &$84\,140$       &$2.82$   &$5.236$        	&$3.67$	&${\rm2.49\,x\,10^{49}}$        &${\rm4.97\,x\,10^{47}}$        	&${\rm3.05\,x\,10^{48}}$    &$2.23$    &mrg01 \\ 
		       $40$      &$79\,430$       &$2.71$   &$5.175$           	&$4.21$        	&${\rm1.85\,x\,10^{49}}$        &${\rm1.08\,x\,10^{47}}$        	&${\rm2.53\,x\,10^{48}}$  &$2.15$	 &sch02 \\  
		        $30$      &$73\,960$       &$2.10$  &$5.271$         	&$5.20$	 &${\rm8.27\,x\,10^{48}}$        &${\rm2.09\,x\,10^{45}}$        	&${\rm1.33\,x\,10^{48}}$    &$2.11$    &mrg01\\  
		         $25$      &$70\,800$       &$1.85$   &$5.301$        	&$6.07$	&${\rm5.33\,x\,10^{48}}$        &${\rm5.90\,x\,10^{44}}$        	&${\rm9.52\,x\,10^{47}}$   &$2.09$    &sch02 \\ 
		        $20$      &$65\,310$       &$1.65$  &$5.305$           	 &$7.53$       	&${\rm3.00\,x\,10^{48}}$        &${\rm1.17\,x\,10^{44}}$        	&${\rm6.49\,x\,10^{47}}$  &$2.04$	&mrg01\\  
		       $15$      &$57\,280$       &$1.48$   &$5.273$         	&$10.37$	 &${\rm1.34\,x\,10^{48}}$        &${\rm1.53\,x\,10^{43}}$        	&${\rm4.06\,x\,10^{47}}$    &$1.95$  &mrg01 \\  
		         $12$      &$49\,890$       &$1.42$   &$5.210$        	&$13.81$	 &${\rm6.25\,x\,10^{47}}$        &${\rm2.73\,x\,10^{42}}$        	&${\rm2.85\,x\,10^{47}}$   &$1.85$   &mrg01 \\ 
		           $9$      &$41\,590$      &$1.34$  &$5.135$        	&$21.04$	  &${\rm1.88\,x\,10^{47}}$        &${\rm2.29\,x\,10^{41}}$        	&${\rm1.76\,x\,10^{47}}$   &$1.66$   &mrg01 \\ 						
		\hline	
		\end{tabular}
		\begin{tablenotes}
			\item[a] The values for the mass, temperature and radius are from 
			the stellar models in \cite{Schaerer2002a}, \textit{sch02}, and \cite{Marigo2001}, 
			\textit{mrg01}, as compiled and presented by \cite{Kubat2012}. For the calculation 
			of the stellar lifetime we have used the analytical expression in Table 6 of 
			\cite{Schaerer2002a}, for the case of zero metallicity and no mass-loss stars. 
			\item[b] The values of the photon flux are in units of ${\rm s^{-1}}$. 
			\item[c] Mean Lyman continuum photon energy in units of Rydbergs.
		\end{tablenotes}
	\end{threeparttable}
	\end{center}
\end{table*}

  In our calculations we only consider the stage when the stars reside in the zero age 
main sequence (ZAMS) and we do not account for stellar evolution. 
Several authors have studied the evolution of massive 
Pop III stars \citep[e.g.,][]{Marigo2001,SchaererPello2002,Schaerer2003,Inoue2011,
Kubat2012}. We have explicitly tested that our choice has little impact on  
our main conclusions. 
Due to the lack of metals, the effect of stellar winds for Pop III stars has been considered to
be not important since the radiation pressure is thought to be very small \citep[e.g.,][]
{Marigo2001,Kudritzki2002,Marigo2003,Schaerer2002a,Krticka2006,Krticka2009}. 
We therefore ignore stellar winds or mass ejections in the computations but we note that other 
authors are revisiting these processes, e.g.,\cite{Vink2015}. We construct only hydrogen and 
helium non-rotating stars. 
Rotation may induce a strong metal mixing from the inner to the outer 
stellar layers, as well as ejection of material to the surrounding medium. This can strongly 
affect the predicted spectra during the evolution of the star, but these effects are expected 
to be smaller during the ZAMS. In addition, the effects of rotation depend on parameters, 
such as velocity and angular momentum, which are still not fully understood for the case 
of Pop III stars \citep[e.g.,][see also \cite{Maeder2012} for a review]{Ekstrom2008,
Yoon2012,Yoon2014,Lau2014}. We assume an helium mass 
fraction of $Y_{He}=0.24$, which implies a fraction in number of $0.06$,
although we have tested that a number fraction of $0.10$ does not produce significant 
differences.

  We use stellar parameters for the mass range $9 -1000\, {\rm M_{\odot}}$ from 
\cite{Kubat2012}, who adopted values for the stellar mass, temperature and radius from the 
works of \cite{Schaerer2002a} and \cite{Marigo2001}. For the calculation of the stellar lifetime 
we have used the analytical expression in Table 6 of \cite{Schaerer2002a}, for the case of 
zero metallicity and no mass-loss stars. The values of these parameters  are shown in Table 
\ref{ta:starparams}, where 
we also show the computed values for Q(HI), Q(HeII), Q(LW) and $\overline{{\rm E}}$. These  
parameters denote the ionizing photon flux for hydrogen, singly-ionized helium, 
the photon flux in 
the Lyman Werner (LW; hereafter) band\footnote{The LW band corresponds to the energy 
range between $11.2\,-\,13.6\,{\rm eV}$, where the photons of such energies are able to 
photo-dissociate the hydrogen molecule ${\rm H_2}$.} and the mean Lyman continuum photon
energy in units of Rydbergs, respectively. 
Our photon flux values are in agreement with those by 
\cite{Schaerer2002a}\footnote{Only our calculations of Q(LW) differ significantly from those 
of \cite{Schaerer2002a}. 
We traced this difference back to the adopted energy range for the LW band. We find 
agreement with the results by \cite{Schaerer2002a} 
using the range $11.2-\infty$. In practice, the upper limit corresponds to the maximum 
frequency value of the stellar SEDs, $5\times10^{16}\,{\rm Hz}$. 
Small variations around this value do not alter the photon fluxes significantly.}.
We stress that the variation of the photon flux in the whole mass range for the case 
of He II (a factor $\sim 10^9$) is much larger than those for hydrogen and the LW band
(factors $\sim 10^4$ and $\sim 10^3$, respectively). This implies 
that the ionized helium emission is much more sensitive to the hardness of the spectra than that 
 from hydrogen, as already noticed by \cite{Schaerer2002a,Schaerer2003}.

\subsection{Stellar population models}\label{sec:pops}

  We compute a set of stellar populations considering different mass ranges 
and power law indexes for an IMF of the form 
\begin{equation}
\xi(m)\,{\rm d}m=m^{-\alpha}\,{\rm d}m ~,
\end{equation} 
where $\xi(m)$ gives the number of stars with mass in the range $m\pm\,{\rm 
d}m/2$, and $\alpha$ is the power law index considered. We use several 
upper mass limits and the values $\alpha=0$ and $\alpha=2.35$, 
which denote a top-heavy and Salpeter slope distributions, respectively. 
A detailed view of the parameters is presented in Table \ref{ta:imfparams}. The
name of each model denotes the IMF parameters, e.g., m9M50a2 meaning a population 
with lower and upper stellar mass limits of 9 and 50 M$_{\odot}$, respectively, and 
Salpeter slope \footnote{All our stellar and galaxy SEDs are publicly available 
at \\ \url{https://github.com/lluism/seds}}. We also show in the table the 
values of the mean Lyman continuum photon energy, in units of Rydbergs, wich will 
be important in following sections. 
\begin{table*}
	\begin{center}
	\caption{IMF model parameters \label{ta:imfparams}}	
	\begin{threeparttable}
		\begin{tabular}{cccccccc} 
		\hline
		\hline
		${\rm Model}$        &$M_{min}\tnote{a}$       &$M_{max}\tnote{a}$        &$\alpha$      &${\rm Q\,(HI)}\tnote{b}$	   &${\rm Q\,(HeII)}\tnote{b}$        &${\rm Q\,(LW)}\tnote{b}$    &$\overline{{\rm E}}$\tnote{c} \\ 
		\hline  
		${\rm m9M50a0}$         &$9$     			&$50$       				 &$0$     		&$3.69$        &$0.04$        	 &$0.52$	    &$2.15$ \\  
		${\rm m9M50a2}$       &$9$      			&$50$        				&$2.35$        	&$1.67$        &$0.01$   		&$0.34$        &$2.04$\\  
		${\rm m9M100a0}$       &$9$        		&$100$       				 &$0$        	&$6.49$        &$0.44$  		&$0.68$        &$2.35$\\  
		${\rm m9M100a2}$         &$9$        		&$100$       				 &$2.35$        	&$2.56$        &$0.09$  		&$0.40$        &$2.11$\\ 
		${\rm m9M500a0}$         &$9$        		 &$500$       				 &$0$           	&$13.53$        &$2.74$   		&$1.00$        &$2.82$\\  
		${\rm m9M500a2}$         &$9$      		&$500$        				&$2.35$         	&$4.91$        &$0.57$  	   	&$0.54$        &$2.23$\\  
		${\rm m50M1000a0}$     &$50$      		&$1000$        				&$0$        		&$14.78$        &$3.26$  		&$1.06$	     &$2.90$   \\ 
		${\rm m50M1000a2}$     &$50$     		&$1000$        				&$2.35$           	&$10.20$        &$1.54$    		&$0.87$        &$2.48$ \\  
		\hline
		\end{tabular}
		\begin{tablenotes}
			\item[a] The values of the lower and upper mass limits are in units of ${\rm M_{\odot}}$. 
			\item[b] The values of the photon flux are in units of ${\rm 10^{47}\,s^{-1}\,M_{\odot}^{-1}}$. 
			\item[c] Mean Lyman continuum photon energy computed by \textit{Cloudy} in units of Rydbergs. 
		\end{tablenotes}		
	\end{threeparttable}
	\end{center}
\end{table*}

\subsubsection{Stochastic sampling of the IMF}\label{sec:stochmethod}

  The stochastic sampling of the IMF has been shown to have a very important effect in cases 
where the deterministic IMF only allows a low number of massive stars (e.g., Salpeter slopes 
instead of flat IMFs), and in cases with low star formation rate or starbursts with 
a small total stellar mass \citep[e.g.,][]{Cervino2000, Bruzual2002,Cervino2006, 
Haas2010,Fumagalli2011,Forero-Romero2013,DaSilva2014}. For simplicity we only 
consider here the sampling of the IMF   
although a more detailed stochastic treatment might also consider the sampling of stellar 
clustering\footnote{Assuming clustered star formation, a fraction of the total stellar 
mass can be divided in smaller parts (the clusters) which are more sensitive to stochastic 
effects \citep{DaSilva2012}.} 
\citep[see, e.g.,][]{Fumagalli2011,Clark2011b,Eldridge2011,Greif2011,DaSilva2012} and stellar 
evolution. For the case of non metal free populations there are codes specially designed 
for such purposes, e.g., the publicly available software \texttt{SLUG} 
\citep{DaSilva2012,Krumholz2015}. 

  Here we ignore the populations with an upper mass limit of 1000 M$_{\odot}$ 
due to the large stellar `mass gap' between 500 and 1000 M$_{\odot}$ in our stellar models. 
To obtain the stochastic stellar populations, we proceed as follows:
We draw stars from our IMFs using the inverse cumulative distribution function   
and we follow the same procedure as in the \texttt{STOP-NEAREST}\footnote{This method 
makes the stochastic solution to converge 
towards the deterministic one for large enough target mass values \citep{Krumholz2015}.}
method in \texttt{SLUG} \citep{Krumholz2015} for sampling the total stellar mass 
(target mass; hereafter). Specifically, this method decides whether or not to 
include the star for which the target mass is exceeded, based on the absolute value of the difference 
between the target and the total masses; the star is included in the sample if the difference 
between the two masses when including the star is less than that without it. We consider 
a default target mass of 1000 M$_{\odot}$ for each starburst but we also 
examine the values 100 and 10\,000 M$_{\odot}$ (we compare these numbers with the 
results from simulations in Section \ref{sec:discussion}). 

  We assess the following two cases: 

\begin{itemize}[leftmargin=0pt,itemindent=20pt]
\item {\it ZAMS phase}: We compute different samples, considering different IMF 
parameters and target masses, containing 10\,000 stochastic stellar populations each. We do not 
consider the effect of time, i.e., the populations are assumed to be at the ZAMS and do not 
evolve. We use our stellar sample in Table \ref{ta:starparams} to compare the results with 
the deterministic calculation. 

\item {\it Temporal evolution}: We repeat the above calculations to obtain samples 
but in this case we allow for any stellar mass within the population, 
which enables smoother evolution profiles of the distributions. We use a cubic interpolation of the 
values in Table \ref{ta:starparams} to obtain the parameters for a given 
stellar mass. We assess the evolution in time in two ways:
First, we study the evolution of galaxies considering only one burst of star formation. 
We compute the distributions at certain time steps considering 
only the stars that are still alive at that time. We repeat this procedure until all the stars in 
the galaxies have disappeared. Second, we follow the evolution of galaxies as before but we 
allow now periodic bursts of star formation. 
 
\end{itemize}

\subsection{Nebular models}\label{sec:nebula}

  In this section, we examine the contribution of the medium around the stellar 
populations to the final spectra. This region will shape the incident stellar radiation field
and will yield a different outcoming spectrum \citep[e.g.,][]{Schaerer2002a, Schaerer2003, 
Osterbrock2006,Schaerer2008,Raiter2010,Inoue2010, Inoue2011,Schaerer2014}. 

  We use the photoionization code \textit{Cloudy} v13.03 \citep{Ferland2013}, which 
considers all relevant physical processes and allows us to avoid the use of, e.g., the usual 
case-B recombination assumption. Case-B is often used in calculations of HII regions but 
\cite{Raiter2010} showed that significant departures from case-B can occur in very low 
metallicity environments. We will examine the case-B recombination assumption for our 
populations in Section \ref{sec:caseb}.

  It is common to use black body spectra to characterize the stellar spectra incident on the 
nebula. However, black body spectrum does not reproduce the stellar spectra 
accurately and several corrections need to be done \citep[e.g.,][]{Tumlinson2000,Rauch2003,
Raiter2010,Schaerer2014}. Therefore, we use our previously computed stellar population SEDs. 

  We assume, following \cite{Schaerer2002a, Raiter2010}, that the 
escape fraction of ionizing photons is $f_{esc}=0$, and that the nebula surrounding the 
stellar population is a static, ionization bounded, spherically closed geometry with a 
constant density. If the escape fraction of ionizing photons is small,  
the variations to the luminosity and photon flux will also be small, almost null for the helium 
line flux \citep{Schaerer2002a}. If the escape fraction is large, the calculations 
rapidly become much more complicated \citep[see, e.g,][]{Inoue2011}, and taking into 
account proper radiative transfer codes and/or numerical simulations for the environment may 
be necessary. Given that the actual value of the escape 
fraction for Pop III populations is not well know, we do not explore more complex 
scenarios in this work. We explore a grid of 
models with different nebular parameters such as density, metallicity and
ionization parameter, covering the parameter space used by \cite{Raiter2010}, who 
explored usual values found in HII regions. The values of these 
parameters are described in Table \ref{ta:nebulaparams}.

\begin{table}
	\begin{center}
	\caption{Values of the parameter grid for the nebula model\label{ta:nebulaparams}}	
	\begin{threeparttable}
		\begin{tabular}{ccc} 
		\hline
		\hline
		$\bf{\rm Parameter}$        &$\bf{\rm Values }$       &$\bf{\rm Units}$               \\ 
		\hline  
		${\rm log\, r_{in}}\tnote{\,a}$	&$17$     					&${\rm log \,(cm)}$       				     \\  
		${\rm log\, U}\tnote{\,b}$         	&$[-4\,, -1]$     			        &$-$       				     \\  
		${\rm log\,n_{H}}\tnote{c\,}$         &$[1,\,2,\,3]$     			&${\rm log\,(cm^{-3})}$       				     \\  
		${\rm log\, Z/Z_{\odot}}\tnote{\,d}$         	&$[0,\,-3,\,-6,\,-9]$     		&$-$       				     \\  
		${f_{esc}}\tnote{\,e}$             	&$0$     					&$-$       				     \\  
		${\rm Spherical\,\, geometry}$         &$ $     			&$ $       				     \\  
		${\rm ionization\,\, bounded}$         &$ $     			&$ $       				     \\  
		${\rm Static\,\, nebula}$		         &$ $     			&$ $       				     \\  		
		\hline
		\end{tabular}
		\begin{tablenotes}
			\item[a] Inner radius of the nebula. 
			\item[b] Ionization parameter, relating gas density and photon flux as 
			${\rm U}=\frac{\rm Q\,(HI)}{4 \pi r_{\rm in}^2\,n_{\rm H}\,c}$, at 
			the inner radius of the nebula. 
			\item[c] Constant total hydrogen density.
			\item[d] Metallicity, expressed relative to the solar metallicity, where $Z_{\odot}=0.02$. 
			\item[e] Nebular escape fraction of ionizing photons.
		\end{tablenotes}		
	\end{threeparttable}
	\end{center}
\end{table}

\section{Results}\label{sec:results}

  In this section we present our main results, focusing on the differences between 
stellar populations for different IMF parameters in Section 
\ref{sec:spectrum}. We only discuss about the differences for a variety 
of nebular parameters when this is relevant for our work; a detailed analysis 
can be found in \cite{Raiter2010}. In Section 
\ref{sec:caseb} we examine the departures from case-B and their origin, and in 
Section \ref{sec:stoch} we explore the effects of the stochastic sampling of the IMF.


\subsection{Spectrum of Pop III stellar populations }\label{sec:spectrum}

We find that populations with massive stars 
can give rise to significant emission of the Lyman 
series and HeII$\,\lambda1640$ lines, 
due to their large stellar ionizing flux and high mean Lyman continuum  
photon energy (we will discuss about the importance of the mean Lyman continuum photon 
energy in Sections \ref{sec:caseb} and \ref{sec:stoch}). The luminosity in the LW band is 
also enhanced above the stellar values \citep{Schaerer2002a,Johnson2008}.
High nebular hydrogen density and very low metallicities boost this behaviour 
in agreement with the results of \cite{Raiter2010}. For the case of populations containing 
only stars of mass ${\rm \le 50\,M_{\odot}}$,  
the LW and Lyman series line fluxes are fainter. More importantly, the HeII$\,\lambda1640$ 
line is dramatically reduced, and its visibility over the continuum strongly depends on the 
metallicity and density of the nebula.

\subsection{Case-B departures}\label{sec:caseb}

\begin{figure*}
\begin{center}
\includegraphics[width=0.5\textwidth]{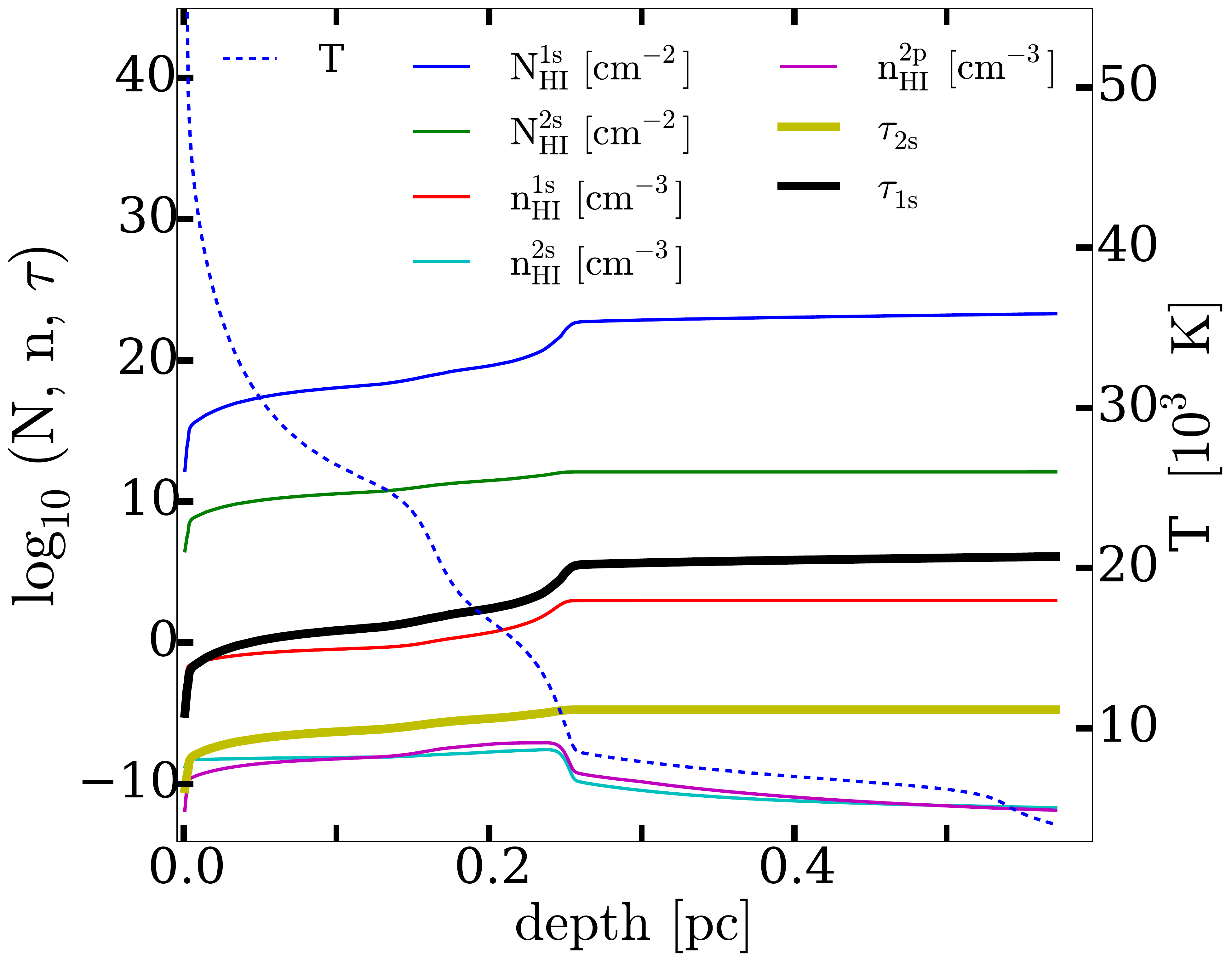}\includegraphics[width=0.47\textwidth]{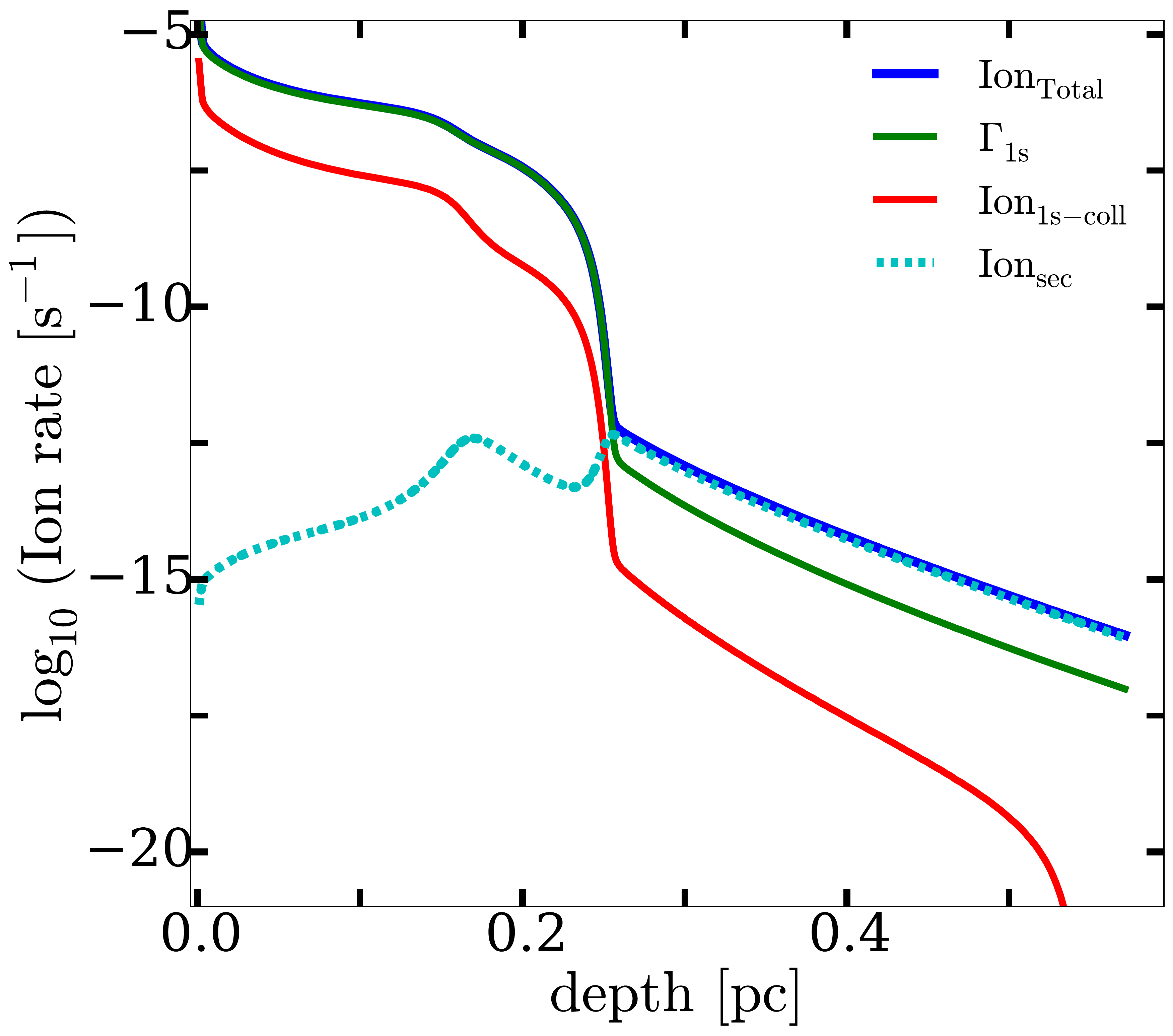}
\caption{Several properties of the nebula surrounding the population
${\rm m50M1000a0}$. The parameters used are ${\rm log\,U=-1}$,  
$\rm n_{H}=10^3\,cm^{-3}$ and ${\rm log\,Z/Z_{\odot} =-9}$. The horizontal axis denotes the depth, 
where the null value is the internal part facing the stellar population. \textit{Left panel:} 
The vertical left axis represents the decimal logarithm of the column density, number density 
and optical depths, with the units shown in the legend. We show the column density (dark blue and 
green lines) and optical depth (thick lines) for the ground state 
and first excited state, and the population of the ground state and the two excited states, 2s 
and 2p, (red and light blue lines) of hydrogen. The vertical right axis denotes the electron temperature (dashed blue line).
The region at $r\sim0.25$ pc denotes the transition from mostly ionized to mostly neutral 
hydrogen. This happens at a T$_e\sim10^4$ K. Note the low values for the column density and 
optical depth of the first excited state compared to those for the ground state, which makes the 
photoionization from the former very unlikely. Note also the high values of the temperature in the 
ionized region which will be the responsible for collisional excitation of the Ly$\alpha$ line by free 
electrons. \textit{Right panel} shows the major contributors to the hydrogen ionization in the nebula. 
the thin green line denotes the photoionization from the ground state which dominates in 
the ionized region ($\sim90\%$ of the total). The collisional ionizations from 
the ground state (red line) represent $\sim10\%$ of the total. In the neutral region, the 
secondary ionizations (dashed blue line) dominate and the photoionization represents $\sim10-
20\%$ of the total.}
\label{fig:caseB}
\end{center}
\end{figure*}

  In this section we first study the existence of departures from the case-B 
assumption in our populations, shown to be important in metal-free objects by 
\cite{Raiter2010}, and second, we revisit the origin of these departures.

  We compute the luminosity for the Ly$\alpha$ and HeII$\, \lambda1640$ lines assuming 
case-B recombination, and compare them with the output from \textit{Cloudy},  
always for our lowest metallicity value, ${\rm log\,Z/Z_{\odot} =-9}$, since we expect the differences 
to be the largest in this case as shown by \cite{Raiter2010}. For the same reason, we make 
use of the populations ${\rm m50M1000a0}$ and ${\rm m9M50a2}$, 
which present the two most extreme cases of stellar populations. 

For the analytical case-B, we follow the assumptions and formalism in \cite{Raiter2010} and 
we adopt for the nebula an electron density $\rm n_{e}=10^2\,cm^{-3}$ and an electron 
temperature $\rm T_{e}=30\,000\,K$. This implies a conversion of ionizing photons into 
Ly$\alpha$ of $\sim 68\%$, and yields to the formulae \citep{Schaerer2002a}  
 
\begin{equation}\label{eq:b}
L_{\rm \alpha}^{\rm B} = 1.04 \times 10^{-11}\, {\rm Q(HI)} ~,
\end{equation}
\begin{equation}
L_{\rm 1640}^{\rm B} = 5.67 \times10^{-12}\, {\rm Q(HeII)} ~,
\end{equation}
where ${\rm Q(HI)}$ and ${\rm Q(HeII)}$ are here the photon flux   
computed by \textit{Cloudy} for comparative reasons.  

  The most important results for our further discussion can be summarized as:
(\textit{i}) case-B overestimates the values for the luminosity of the HeII$\, \lambda1640$ 
line by a factor $\sim2$ for the case of high nebular densities, $\rm n_{H}=10^2-10^3\,cm^{-3}$, 
and low ionization parameter, ${\rm log\,U=-4}$, for both populations. 
(\textit{ii}) The values for the Ly$\alpha$ luminosity are underestimated by a factor $\sim3$ 
for our high-mass stars population and $\sim2$ for the low-mass stars one. These values 
closely match the mean Lyman continuum photon 
energy in units of 1 Rydberg \citep[as noted previously by][]{Raiter2010}. For high densities, 
$\rm n_{H}=10^2-10^3\,cm^{-3}$, the departures are slightly higher than for lower densities.
These results are in very good agreement with those obtained originally by \cite{Raiter2010}. 

As an explanation for the case-B departure of the Ly$\alpha$ radiation, 
\cite{Raiter2010} proposed that the 
high temperatures in the ionized region of the nebula allow for the collisional 
excitations to populate the first excited state of the hydrogen atom. Then, 
ionization by low energy photons can occur from that level, thus boosting the 
emission of Ly$\alpha$ to larger values compared to case-B (which only accounts for 
photoionizations from the ground state). 
However, for the photoionization from the first excited state to occur, very 
large column densities for this state are necessary since the cross section of this transition is
of the order $\sigma_{n=2}\sim10^{-17}\,{\rm cm^{-2}}$. Having a non-negligible optical 
depth to photoionization from the $n=2$ state may imply that the nebular region becomes 
optically thick in the HI fine structure lines \citep[see ][]{Dijkstra2016}

  In order to revisit the processes governing the departures from case-B, we use   
the population ${\rm m50M1000a0}$. The parameters for the nebula are
${\rm log\,U=-1}$, $\rm n_{H}=10^3\,cm^{-3}$ and ${\rm log\,Z/Z_{\odot} =-9}$. This set 
of parameters is the one favouring the largest differences between \textit{Cloudy} and 
the analytical calculation. For this population, the mean Lyman continuum photon energy is $2.9$
Ryd. 

  The {\it left panel} in Figure \ref{fig:caseB} shows several parameters of the nebula as 
a function of depth (x-axis), where $r$ denotes the distance into the nebula (which begins 
at $r_{in}=10^{17}\,{\rm cm}$). Two regions besides $\sim0.25$ pc are clearly visible, 
denoting the mostly ionized (left part) and mostly neutral (right part) nebular hydrogen
regions. The populations for the 2s and 2p excited states ({\it light blue} and {\it purple thin lines}, 
respectively) appear to be $\sim 9$ orders of magnitude below that of the ground state 
({\it red thin line}) in the ionized region, and the difference is larger in the neutral part of the nebula. 
This difference is the same for the column densities ({\it green} and {\it dark blue lines}) and  
similar for the optical depth ({\it yellow} and {\it black thick lines}), due to differences in the 
cross-section of the ground and first excited states. This implies that the first excited state 
is very optically thin to ionizing radiation.
The {\it right panel} denotes the processes contributing to the hydrogen ionization from {\it Cloudy}.
In the ionized part, the electron temperature reaches very high values allowing for the 
collisional processes of ionization from the ground state to be important 
({\it red line}), representing $\sim10-15\%$ of the total ionization. The other $\sim 90\%$ 
is due to photoionization from the ground state ({\it green solid line}). In the neutral part, 
secondary ionizations are the major mechanism ({\it dotted light blue line}), although the 
total ionization rate is now many orders of magnitude lower. In any case, 
photoionizations from the first excited state are negligible. 
{\it Cloudy} shows that $\sim 40\%$ of the Ly$\alpha$ flux is 
due to collisionally excited Ly$\alpha$ cooling radiation. Therefore, 
we conclude that Ly$\alpha$ departures from case-B  arise mainly due to 
collisional excitation and ionization of hydrogen atoms by energetic free electrons.
In addition, there is contribution 
to the Ly$\alpha$ production from the collisional mixing between the 2s and 2p levels for the 
case of high density regions \citep{Raiter2010}, helium recombinations and secondary ionizations.

\subsection{Stochastic IMF effects}\label{sec:stoch}

  We show the effects of the stochastic sampling of the IMF in the two following sections. 
We first present the results for bursts of star formation at the ZAMS in Section 
\ref{sec:stochzams} and then the results regarding time evolution in Section 
\ref{sec:stochtime}. With this stochasticity study accounting for time 
evolution we aim to provide a basic framework/intuition for 
extending our results to more general star formation histories. 
Our discussion focuses on the default target mass (total stellar mass) 1000 M$_{\odot}$, 
although we also show and discuss the cases with target masses 100 and 10\,000 M$_{\odot}$.

\subsubsection{ZAMS phase}\label{sec:stochzams}

  We present the results for the samples containing 10\,000 galaxies at the ZAMS, 
built using the stochastic IMF method described in Section \ref{sec:stochmethod}. 

\begin{figure}
\begin{center}
\includegraphics[width=0.48\textwidth]{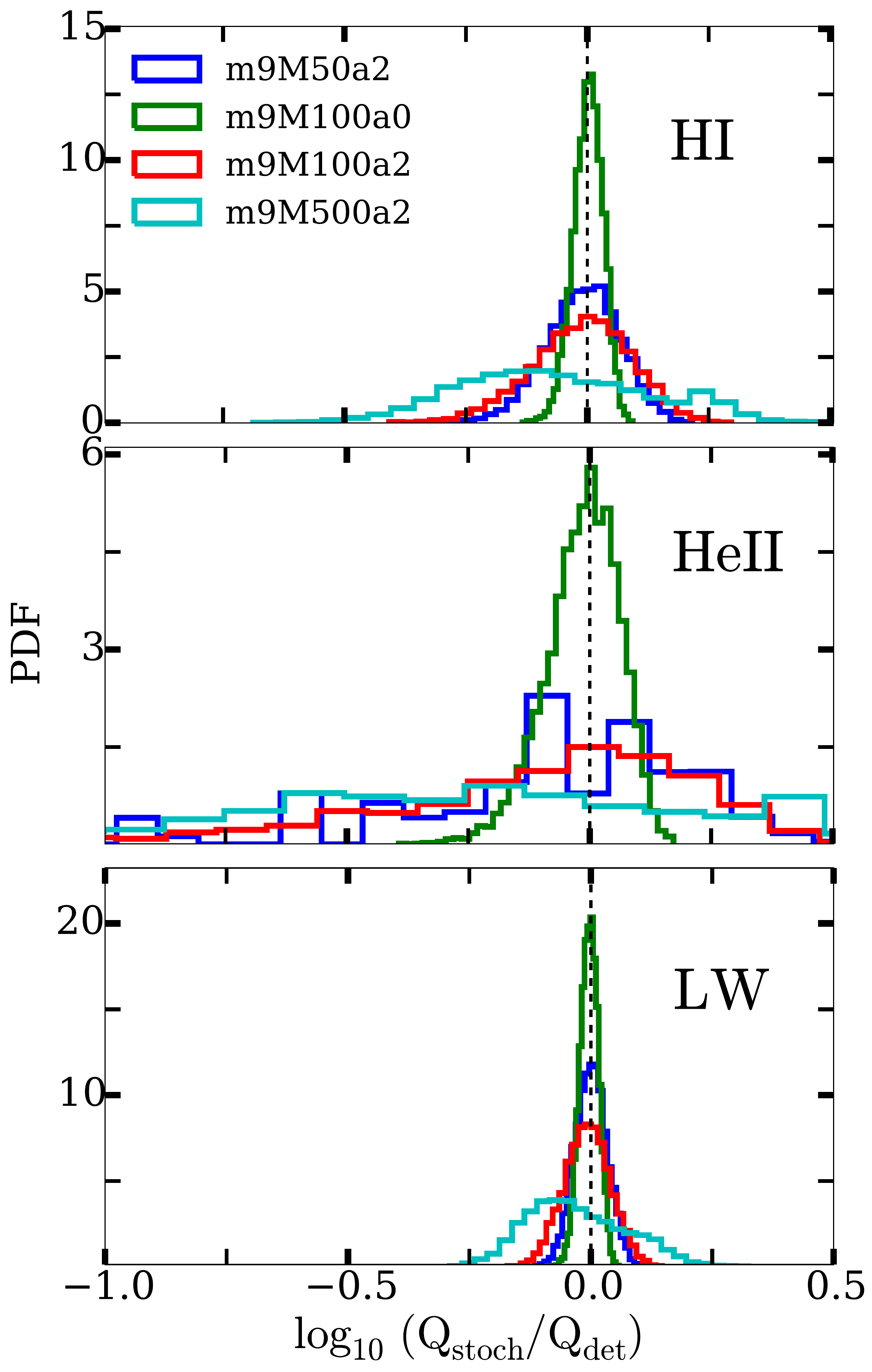}
\caption{Normalized distributions of ionizing and LW photon flux 
for several IMFs relative to the 
deterministic calculations ({\it vertical dashed line}). The target mass value is 1000 M$_\odot$. 
From top to bottom, the distributions are for 
hydrogen, single-ionized helium and LW band. Note the strong stochastic effect 
for IMFs with Salpeter slopes compared to that for flat IMFs. The HeII distribution shows
the broadest shape due to the strong dependence of Q(HeII) on the stellar mass. The stochastic
effect increases with higher upper mass limits. }
\label{fig:Qdistr}
\end{center}
\end{figure}

  Figure \ref{fig:Qdistr} shows the normalized distributions of 
ionizing and LW photon fluxes, relative to the deterministic values 
({\it vertical dashed lines}), for several IMFs. As expected from previous studies, 
the stochastic effects are more important for those populations favoring the presence 
of a low number of massive stars with respect to the total number of stars in a burst, 
i.e., Salpeter slope, and for high upper mass 
limit IMFs. The distributions for helium present the broadest 
shapes ( $>$1 order of magnitude below 
and a factor $\gsim$3 above the deterministic value for all IMFs except for the flat one). 
This is due to the strong dependence on the hardness of the spectra, which in turn, 
depends on the 
stellar mass distribution in the populations. For the case of hydrogen 
and LW, the distributions are much narrower since the photon flux in these cases is less sensitive 
to the stellar mass. The distributions of helium flux for the case of target mass 
100 and 10\,000 M$_\odot$ 
reach a maximum value of $\sim 1$ order of magnitude and less than a factor 2  
above the deterministic values, respectively, denoting the sensitivity of the stochastic effects 
to the target mass. For hydrogen and the LW band, the differences compared to the default 
target mass are smaller but still significant.

\begin{figure*}
\begin{center}
\includegraphics[width=0.49\textwidth]{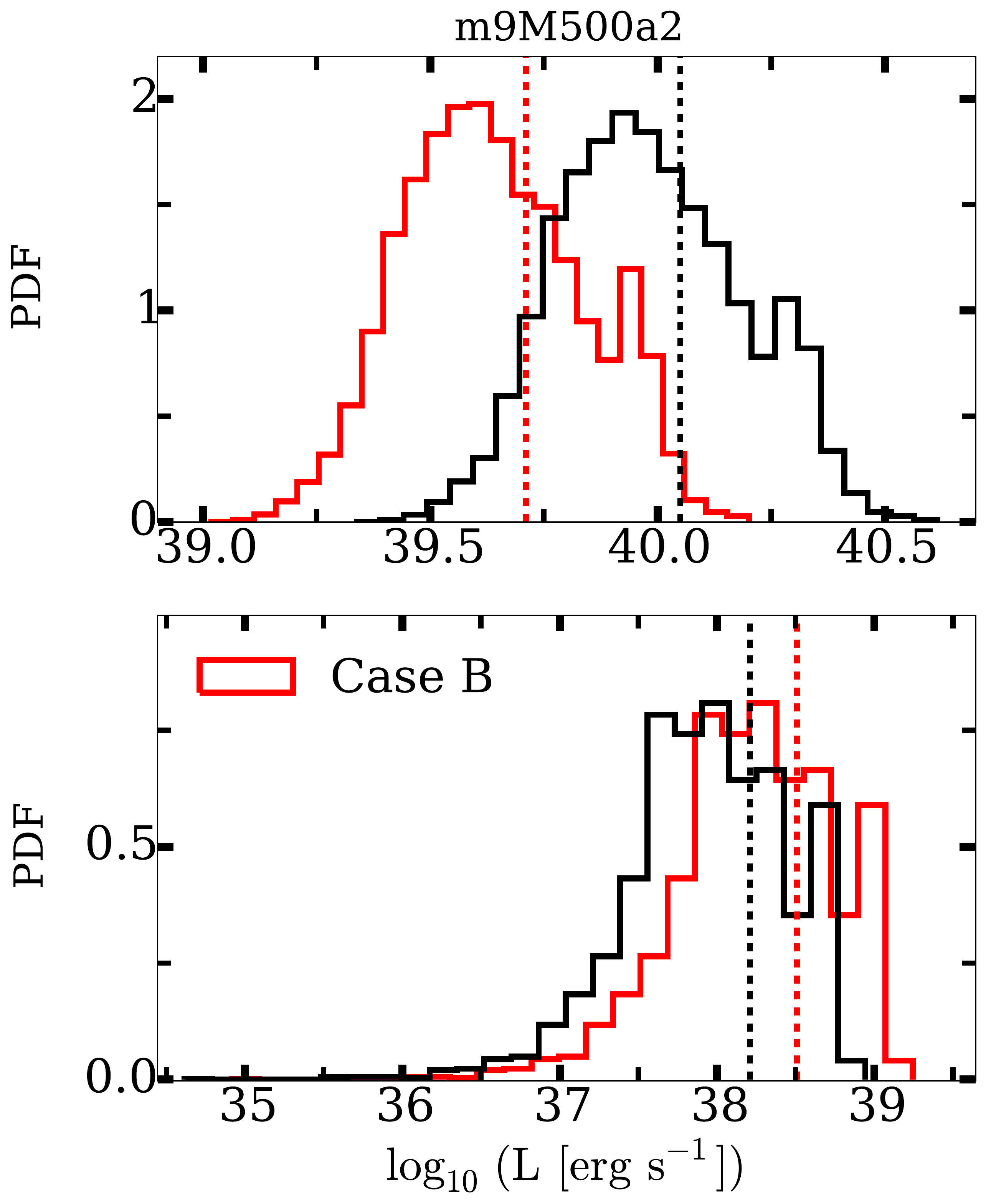}\includegraphics[width=0.45\textwidth]{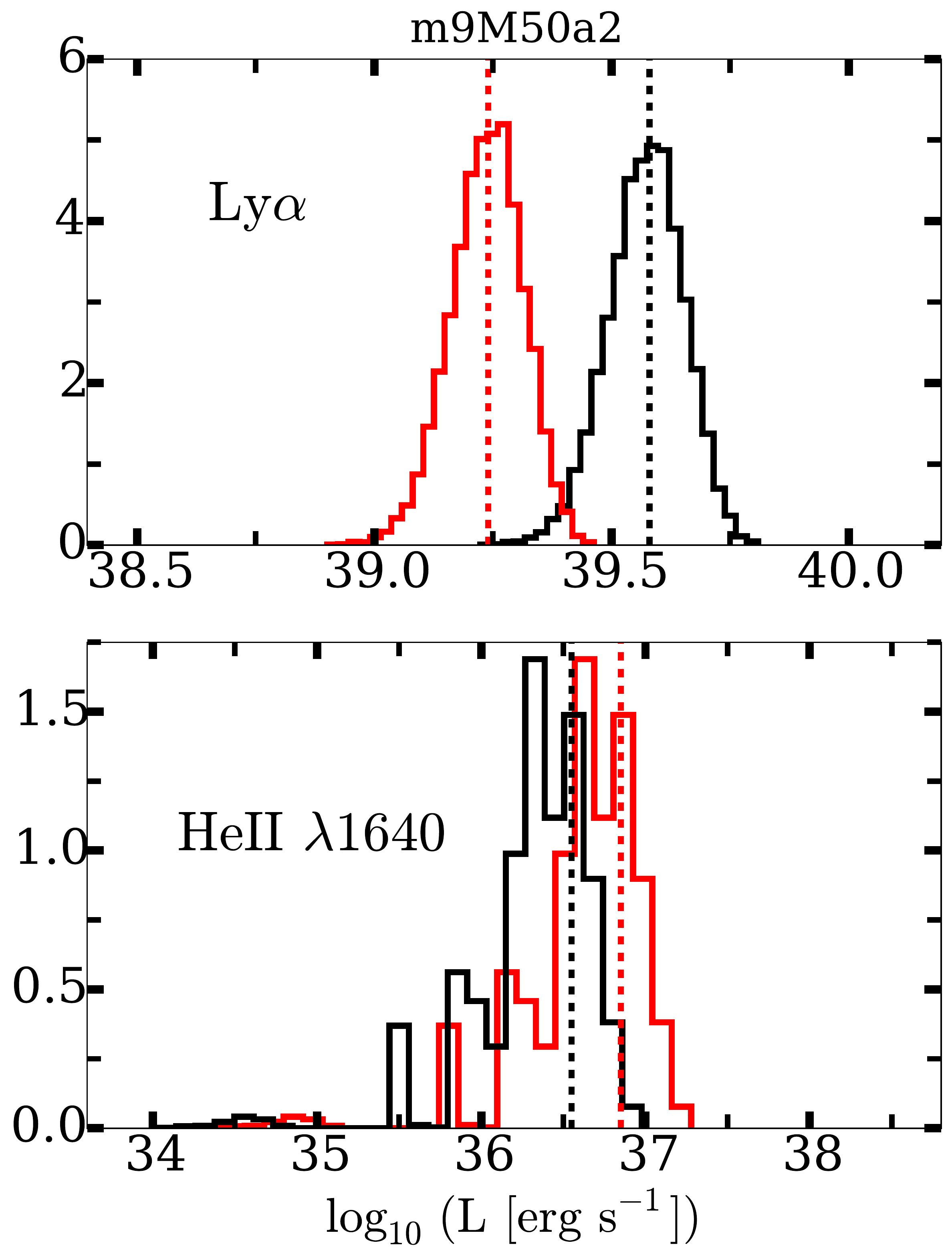}
\caption{{\it Upper panels:} Normalized Ly$\alpha$ luminosity distributions considering  
stochastic effects, assuming case-B ({\it red solid lines}) and not assuming case-B ({\it black solid lines}). 
{\it Lower panels}: Same as before but for HeII$\, \lambda1640$. {\it Left panels} show the cases 
for the IMF ${\rm m9M500a2}$ and {\it right panels} for  ${\rm m9M50a2}$. {\it Vertical 
dashed lines} show the deterministic values. The total target mass is 1000 M$_\odot$.
The effect of stochasticity added to the case-B departure can boost the 
Ly$\alpha$ luminosity a factor $\sim 8$ ($\sim 3.5$) above the common calculations for the case of 
${\rm m9M500a2}$ (${\rm m9M50a2}$).
The effect of case-B overestimating the HeII$\, \lambda1640$ luminosity is due to our
arbitrary selection of the ionization parameter. For other values of this parameter, the differences
between the two distributions are almost null. To facilitate the comparison, the horizontal axis for 
the same transition in the two IMFs has the same length.}
\label{fig:Ldistr}
\end{center}
\end{figure*}

  We showed in Section \ref{sec:caseb} that the departure of the Ly$\alpha$ luminosity from 
the case-B assumption depends on the hardness of the spectra, i.e., on the mean Lyman
continuum photon energy. Therefore, stochasticity can further enhance this difference 
since the hardness of the spectra depends on the distribution of stellar masses. 
To quantify this effect, we calculate the stochastic distributions for Ly$\alpha$ and 
HeII$\, \lambda1640$ luminosities with and without the case-B assumption. We use the IMFs 
${\rm m9M500a2}$ and ${\rm m9M50a2}$ as representative for this calculation 
since they denote the cases where stochasticity has a very high and low effect, respectively. 
For the case-B calculations we adopt the same formalism as in Section \ref{sec:caseb}. 
Due to the high computational cost of running 10\,000 different stellar populations with 
\textit{Cloudy}, for the non-case-B calculations we use the fitting formulae given by \cite{Raiter2010} instead.
This yields a value for the Ly$\alpha$ luminosity
\begin{equation}\label{eq:lyab}
L_{\rm \alpha} = L_{\rm \alpha}^{B} \times \bar{E} \times \frac{\tilde{f}_{\rm coll}}{2/3} ~,
\end{equation}
where $L_{\rm \alpha}^{B}$ is the luminosity assuming case-B (Equation \ref{eq:b}), 
$\bar{E}$ denotes the mean 
Lyman continuum photon energy in Rydbergs and $\tilde{f}_{\rm coll}$ accounts for the 
density effects. For simplicity, we consider a nebular 
density $\rm n_{H}=10^3\,cm^{-3}$ which yields to $\tilde{f}_{\rm coll}=0.78$ following 
equation (9) in \cite{Raiter2010}. Small variations of the density give rise to a few percent variation
in the total result \citep[see section 4.1.1 in][]{Raiter2010}. We assume, again for simplicity, that 
the ionization parameter is ${\rm log\,U=-4}$. This has no effect on Ly$\alpha$  but implies 
that case-B overestimates HeII$ \, \lambda1640$ emission by a factor $\sim2$, due to the 
actual competition between H and HeII for ionizing photons in cases with low U  
\citep{Stasinska1986,Raiter2010}. We will simply 
use this value for computing the non case-B helium luminosity. We 
stress that this is an arbitrary selection and for other values of the ionization parameter and 
same density, the departures from case-B for the case of helium are almost null 
\citep[see Figure 10 in][]{Raiter2010}.

\begin{figure*}
\begin{center}
\includegraphics[width=0.354\textwidth]{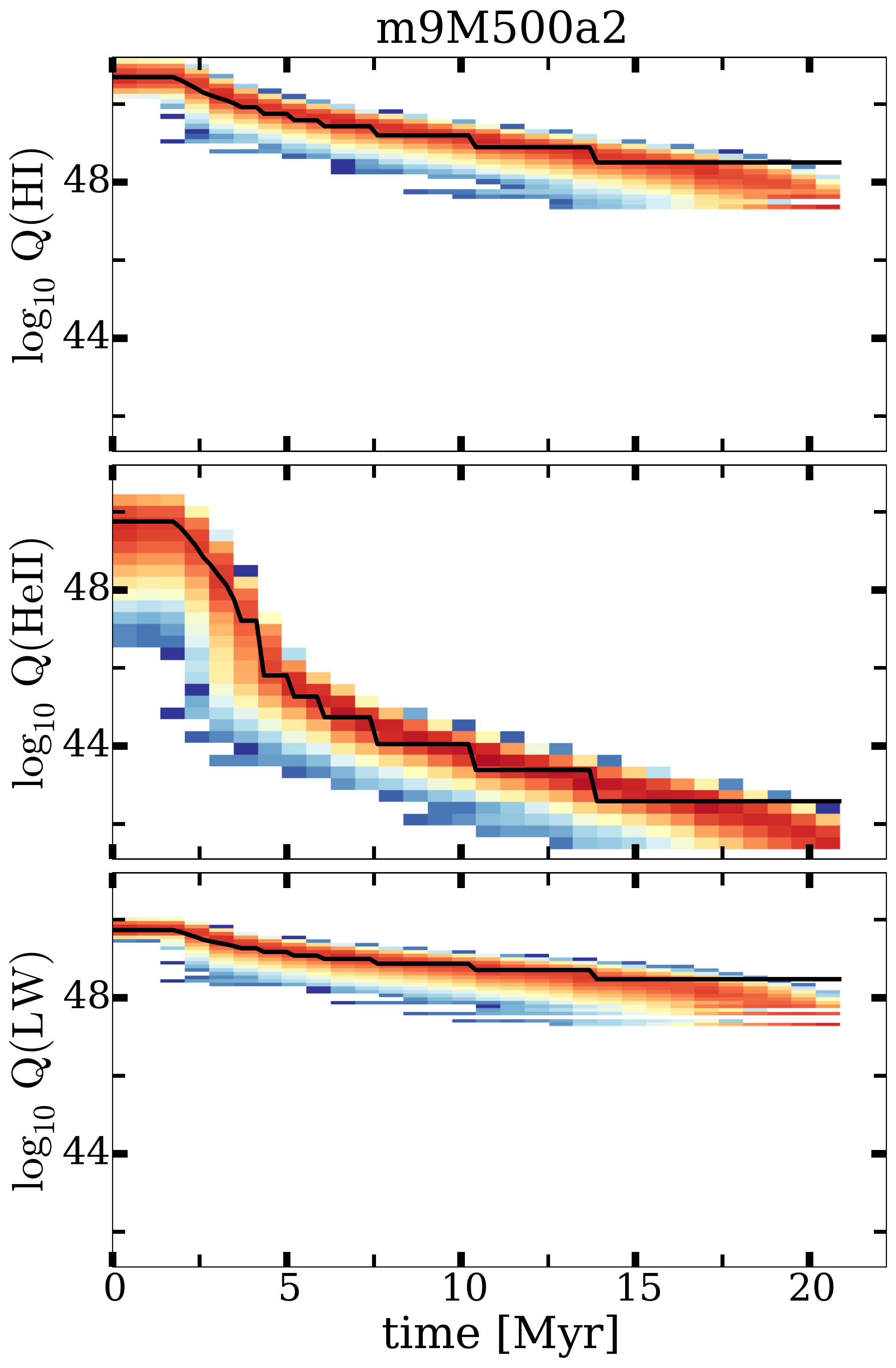}\includegraphics[width=0.317\textwidth]{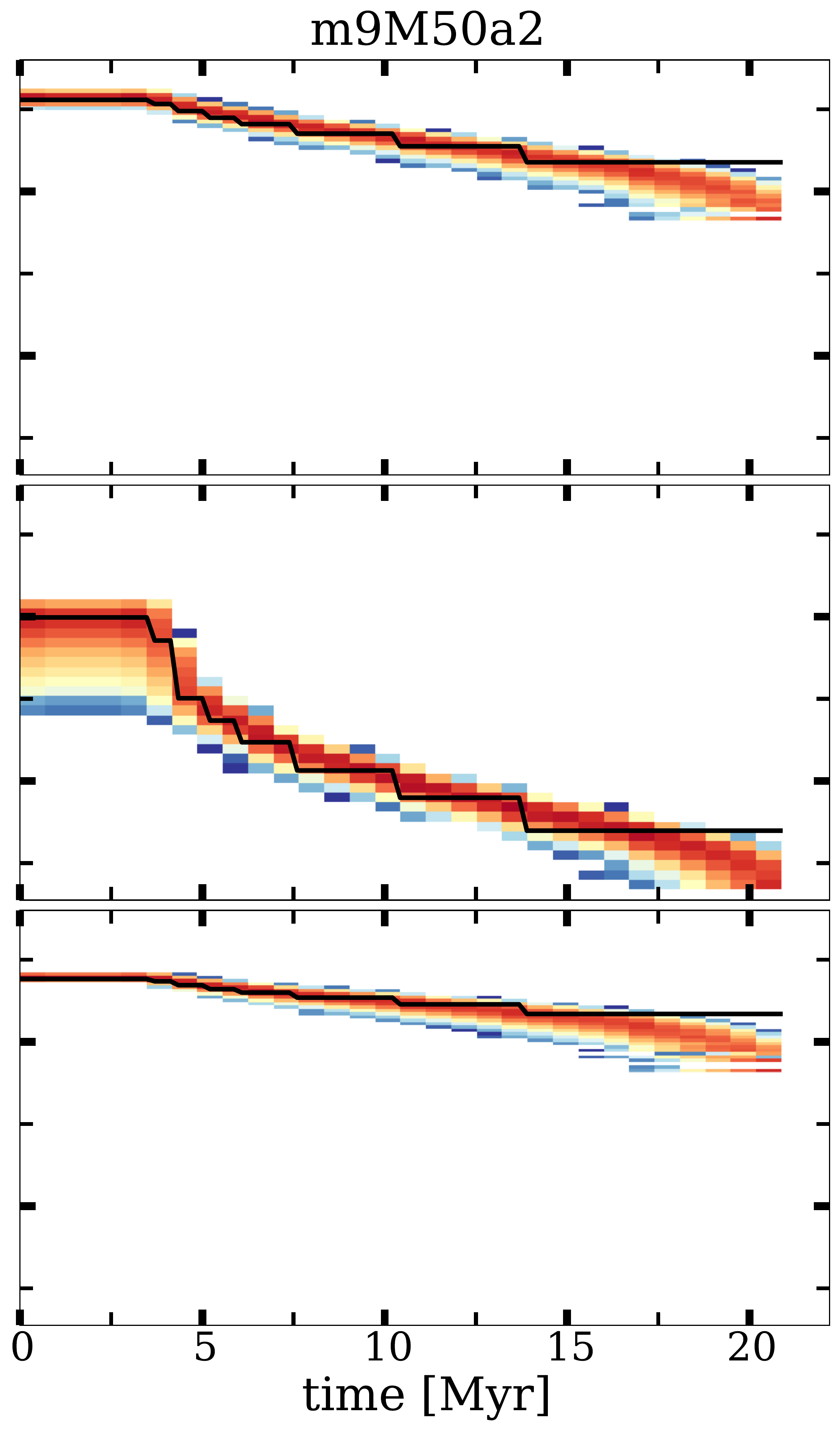}\includegraphics[width=0.33\textwidth]{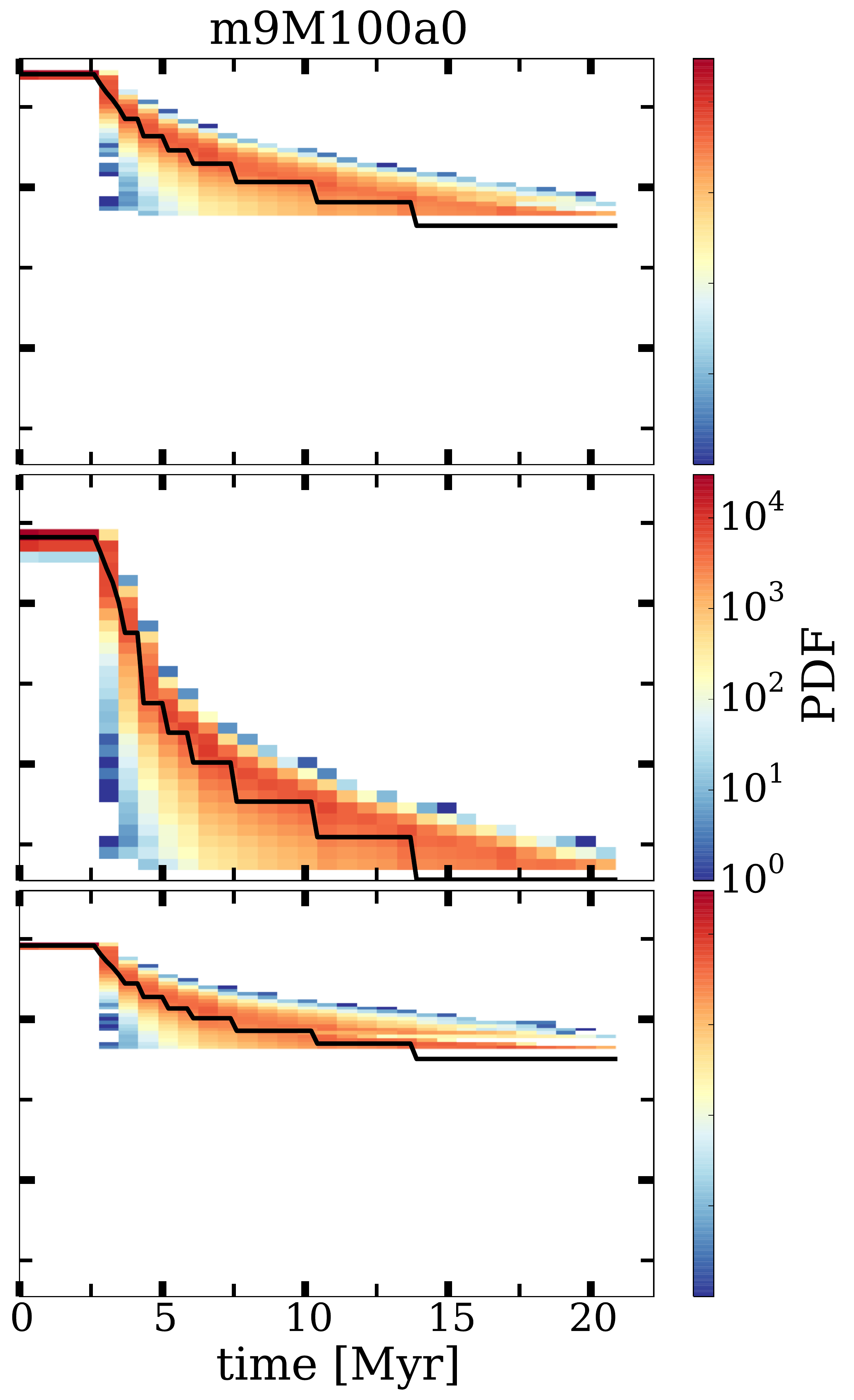}
\caption{Normalized distributions of the evolution of photon flux with respect of time for starbursts 
with IMFs ${\rm m9M500a2}$ ({\it left panel}), ${\rm m9M50a2}$ ({\it central panel}) and 
${\rm m9M100a0}$ ({\it right panel}), from left to right, respectively, and for the default 
target mass 1000 ${\rm M_{\odot}}$. 
{\it Black solid lines} denote the deterministic calculations. For the flat IMF, the distribution 
broadens rapidly when the first stars start to disappear. To facilitate comparison, all the vertical 
axis have the same length.}
\label{fig:decay1000}
\end{center}
\end{figure*}

  Figure \ref{fig:Ldistr} shows the results after these calculations for the default target mass 
1000 M$_\odot$, where the results assuming 
case-B are plotted in red. The {\it vertical dashed lines} show the deterministic values for the 
corresponding distributions. {\it Left panels} present the cases for ${\rm m9M500a2}$ 
and {\it right panels} for ${\rm m9M50a2}$. {\it Upper panels} show the results for 
Ly$\alpha$ and the {\it lower panels} those for HeII$ \, \lambda1640$.  We see that the 
HeII$ \, \lambda1640$ luminosity can be boosted by a factor of $\gsim 3$ compared to the 
deterministic calculation in both populations. The case of Ly$\alpha$ is more interesting 
since the stochastic boost 
(a factor $\sim 3$ for ${\rm m9M500a2}$ and $\sim 1.5$ for ${\rm m9M50a2}$) 
can be added to the departure from case-B (a factor $\sim 2.5$ in both 
cases). The joint effects may give rise up to a factor  
$\sim 8$ and $\sim 3.5$ for the respective IMFs compared to the common calculations 
that do not consider stochasticity and case-B departures. For the 100 M$_\odot$ 
target mass, the total Ly$\alpha$ boost can reach 
a very large value (a factor $\sim 16$ and $\sim 9$ for the two IMFs, respectively) and 
roughly the same values for the case of helium just accounting for the stochastic effects. 
The distributions are much narrower for the case of the largest target mass. In this case, 
the total maximum boosts are hardly above factors of a few in all populations and transitions.

\subsubsection{Temporal evolution}\label{sec:stochtime}

  In this section, we discuss the results for the distribution of photon fluxes 
and luminosities taking into account time evolution for two cases: ({\it i}) A single burst of star 
formation, and ({\it ii}) allowing for periodic bursts. As before, we use the IMFs 
${\rm m9M500a2}$ and ${\rm m9M50a2}$, which have very different sensitivity to 
stochastic effects, and add now ${\rm m9M100a0}$ to investigate the effect of time to flat IMFs.

  {\it First}, in Figure \ref{fig:decay1000} we present the evolution of the distribution of photon 
fluxes considering a single burst of star formation for the default target stellar mass 1000 
${\rm M_{\odot}}$. The {\it black solid lines} show the deterministic calculations, which 
in general follow the peak of the stochastic distributions. The differences at later times 
exist because the stochastic case can only account for an integer number of stars while 
the deterministic approximation can take any real number. This can be 
seen in the last time step of the right panel, where the deterministic number of 9 M$_\odot$ 
stars is lower than unity, so the stochastic distribution can never reach those flux values. 
As expected, the higher upper mass limit IMFs, and specially for the case of helium, 
present the broadest distributions which, in general for all transitions, tend to slightly broaden 
with time. This broadening is much more important for the flat IMF, which can reach the 
minimum value just when the massive stars begin to disappear. This is because in these 
cases the flat IMF  has favored the formation of populations with predominantly massive stars.
For the lowest target mass the distributions are much broader and the 
minimum value of the photon flux can be reached at early times for the same reason as 
explained above. For the largest target stellar mass the distributions are narrow and show 
small differences compared to the deterministic calculations.

\begin{figure*}
\begin{center}
\includegraphics[width=0.3652\textwidth]{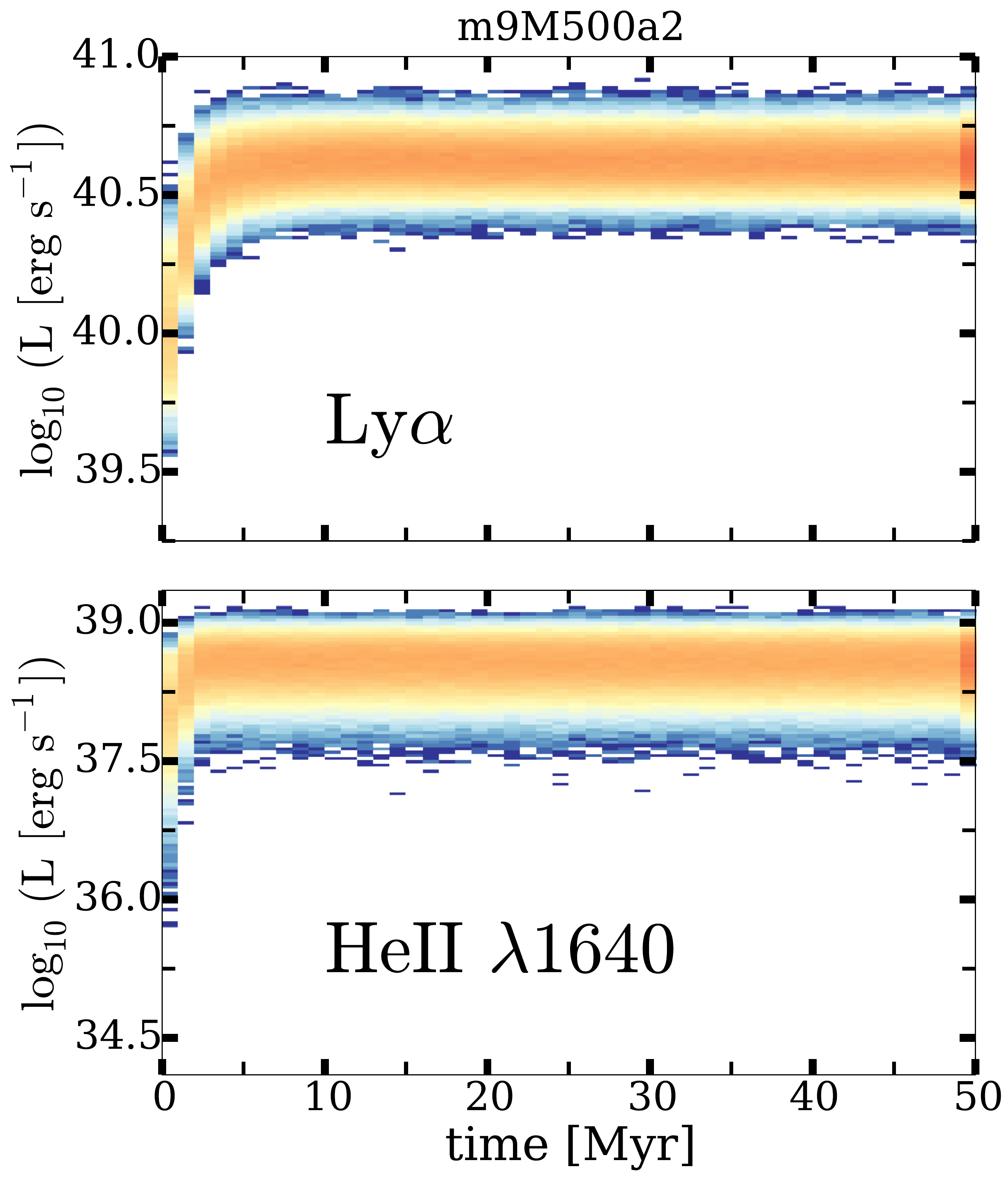}\includegraphics[width=0.315\textwidth]{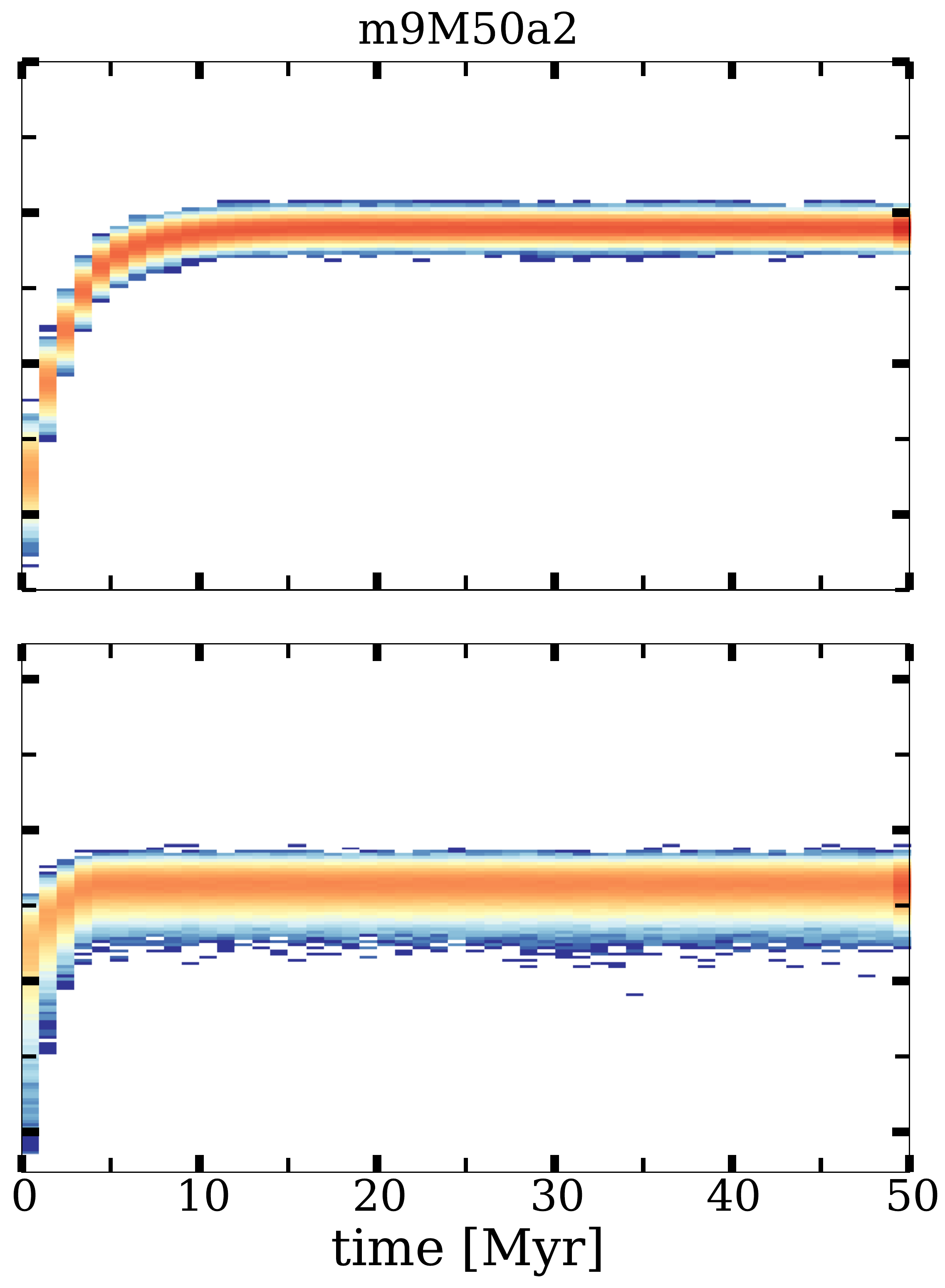}\includegraphics[width=0.315\textwidth]{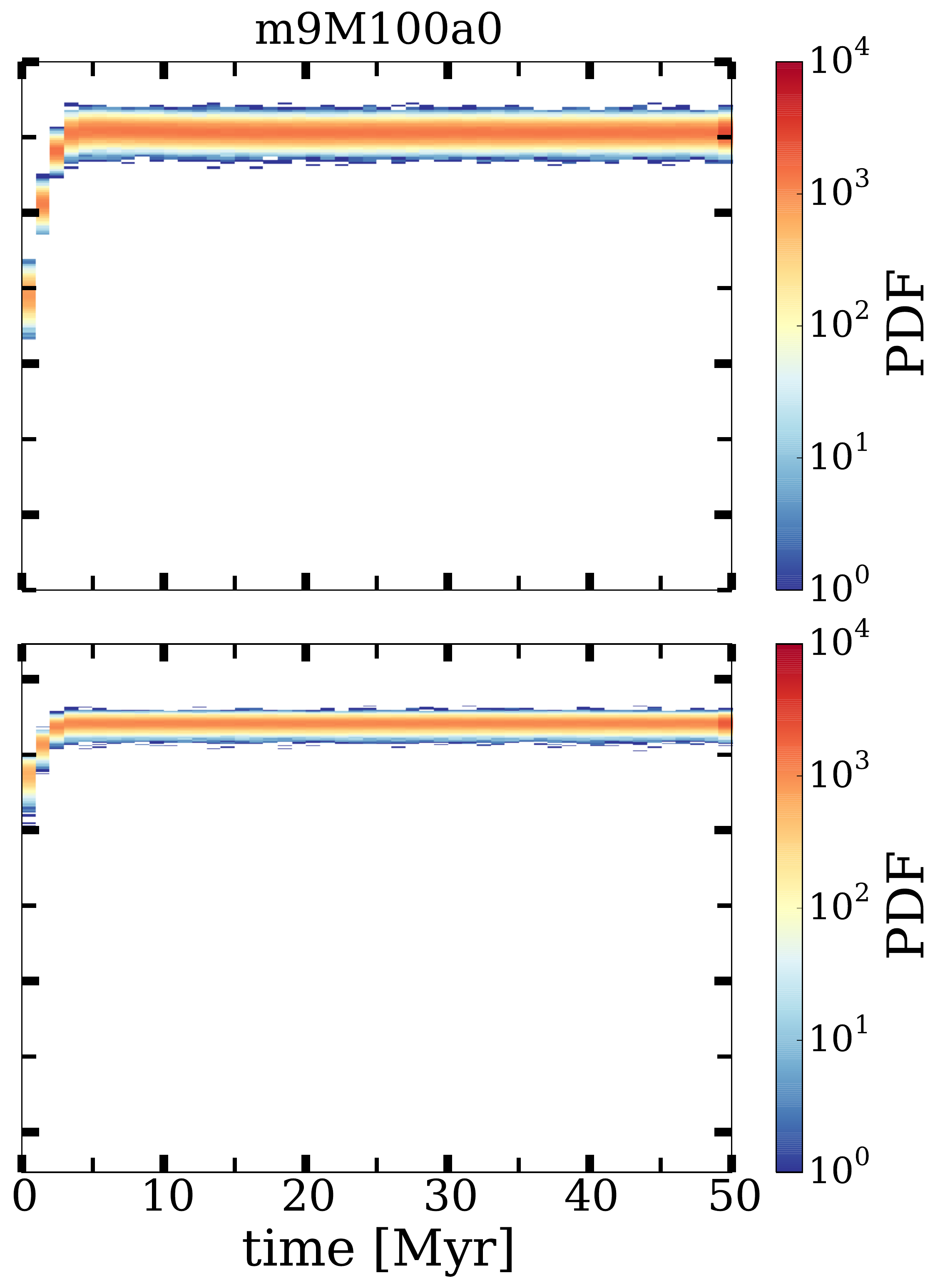}
\caption{Normalized distributions for the evolution of luminosities with respect of 
time, accounting for case-B departures using the formalism shown in Section \ref{sec:stochzams} 
and periodic bursts of star formation of 1000 ${\rm M_{\odot}}$ every Myr. 
From left to right, the IMFs ${\rm m9M500a2}$ ({\it left panel}), ${\rm m9M50a2}$ 
({\it central panel}) and ${\rm m9M100a0}$ ({\it right panel}), respectively.}
\label{fig:csfr1000_Myr_L}
\end{center}
\end{figure*}

\begin{figure*}
\begin{center}
\includegraphics[width=0.47\textwidth]{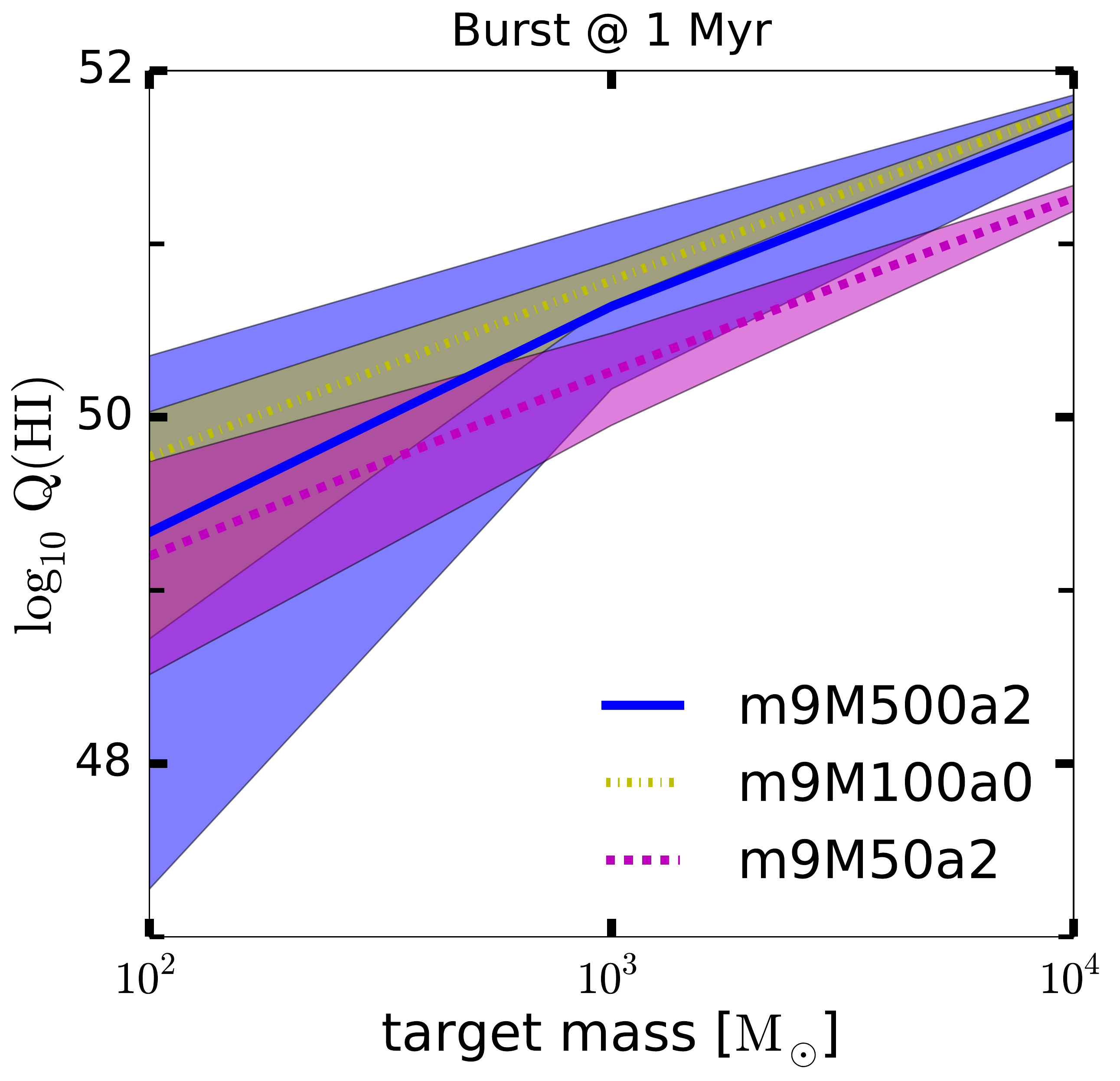}\includegraphics[width=0.425\textwidth]{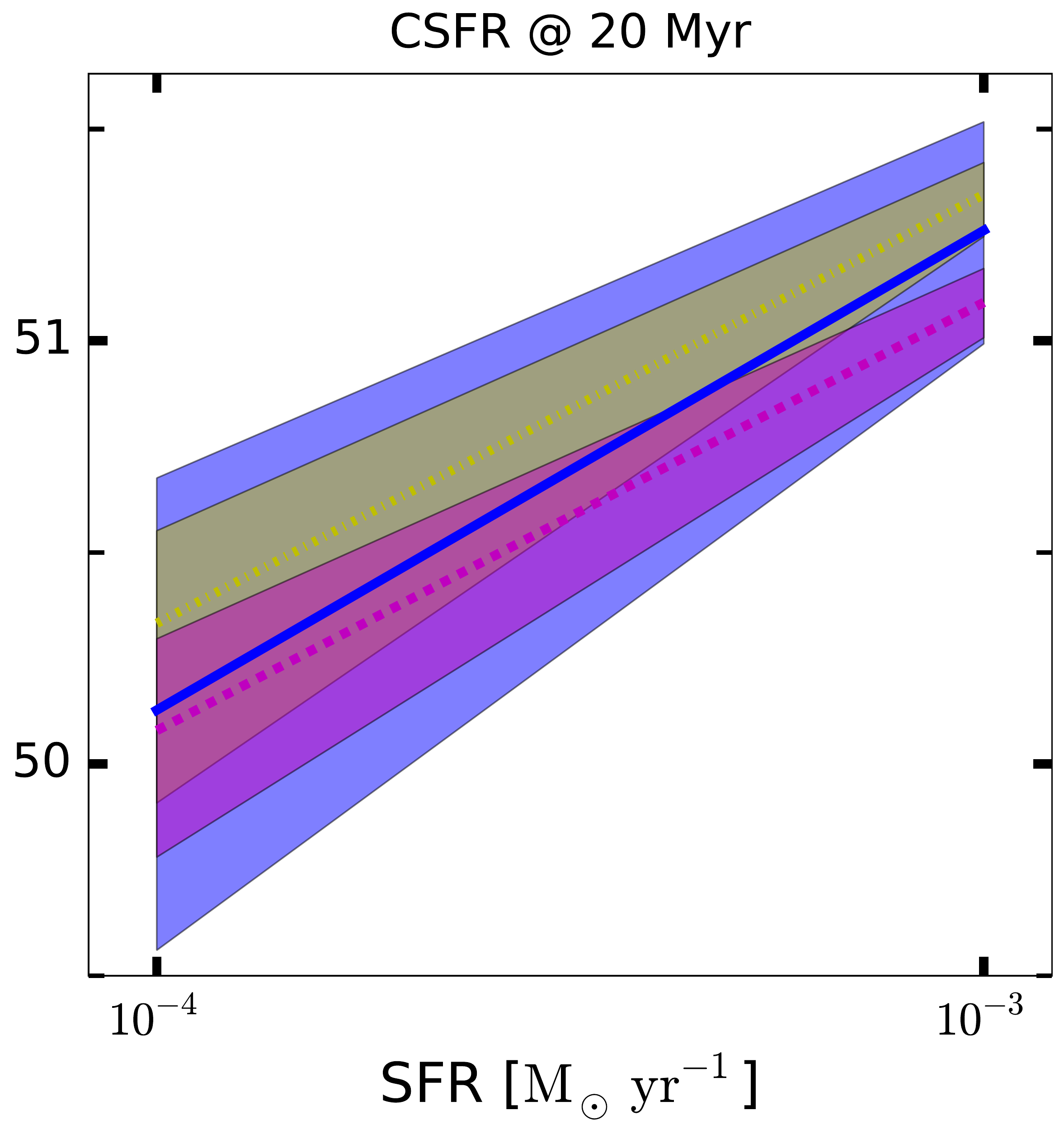}
\caption{Evolution of hydrogen photon fluxes for several populations. {\it Left panel} shows the 
case accounting for a single starburst after 1 Myr for our 3 target masses. Lower target masses 
produce a wider stochastic distribution but cannot compensate for the lower total stellar mass.
{\it Right panel} shows the case for several bursts after 20 Myr to ensure that the photon flux 
has reached the flat plateau seen in Figure \ref{fig:csfr1000_Myr_L}.}
\label{fig:evo}
\end{center}
\end{figure*}

  {\it Second}, we show in Figure \ref{fig:csfr1000_Myr_L} the same above distributions  
but allowing now for periodic bursts of star formation and for the case of Ly$\alpha$ and 
HeII$\, \lambda1640$ luminosities. 
In this case, we create a new stochastic burst using a target mass of 1000 M$_\odot$ every 
Myr, which represents an average star formation rate ${\rm SFR}= 10^{-3}$ M$_\odot \, 
{\rm yr^{-1}}$. We account for Ly$\alpha$ departures from case-B 
adopting the same formalism and values used for Equation \ref{eq:lyab} in Section 
\ref{sec:stochzams} and we assume that there are no departures from case-B for helium.
Interestingly, the Ly$\alpha$ luminosity reaches high values, $L_{\rm \alpha} 
\gsim 5\times10^{40}\,{\rm erg\, s^{-1}}$, for the populations in the left and right panels, and 
a factor of $\sim2-3$ below for the population in the middle panel. 
For the case of helium, the highest upper mass limit population in the left panel reaches 
a maximum value of $L_{\rm 1640}\sim 10^{39}\,{\rm erg\, s^{-1}}$ due to the stochastic 
effects. This plot gives an intuition for the values of the luminosities associated 
with our models.

Our results can be extrapolated to other target masses and SFRs. 
Figure \ref{fig:evo} shows examples of this for a single starburst after 
1 Myr and continuous SFRs after 20 Myr. Note that for the case of a 
single starburst the values in Figure \ref{fig:evo} and the ones for luminosity presented 
above only last for $2-3$ Myr, while the most massive stars are still alive 
(Figure \ref{fig:decay1000}). Figure \ref{fig:evo} shows that when considering 
a target mass of 100 M$_\odot$ 
or ${\rm SFR=10^{-4}}$ M$_\odot \,{\rm yr^{-1}}$, the stochastic effects are larger 
but not enough to compensate for the reduced total stellar mass.
On the other hand, for a target mass of 10\,000 M$_\odot$, the stochastic 
effects are small but the photon fluxes and luminosities are a factor of $\sim 
6-9$ above those for a target mass 1000 M$_\odot$ due to the larger stellar mass. 
This implies luminosities as high as $L_{\rm \alpha}\sim 5\times10^{41}\,
{\rm erg\, s^{-1}}$ and $L_{\rm 1640}\sim 6\times10^{39}\,{\rm erg\, s^{-1}}$. 
We have checked with a simple linear extrapolation of these fluxes that 
for higher SFR values the stochastic broadening disappears quickly after 
SFR$=10^{-3}\,{\rm M_\odot\,  yr^{-1}}$. Only the stochastic effects for 
the IMF m9M500a2 are visible slightly further although never reaching 
the point SFR$=10^{-2}\,{\rm M_\odot\,  yr^{-1}}$.

In the next section, we discuss the factors that may increase or reduce the 
luminosity values that we have found and the possibility for a detection of 
Pop III populations.

\section{Discussion}\label{sec:discussion}

   We discuss in this section the factors that can affect our 
results and the possibility for a detection of Pop III galaxies. 

   Our stellar populations cover the mass range $9-1000\,{\rm M_{\odot}}$. We adopted 
the lower mass limit accounting for the validity of the plane parallel assumption in our
stellar atmosphere modelling. However, we have tested the inclusion of low mass stars 
in our stochastic calculations, simply taking the values of photon fluxes for stars down 
to $1.5\,{\rm M_{\odot}}$ in \cite{Marigo2001}. This implies a broadening of the 
IMFs, i.e., an increase of the stellar mass range. The effect on the 
stochastic distributions is not very significant for the case of hydrogen photon flux and
Ly$\alpha$ luminosities. It is important, however, for the HeII distributions due to their  
strong dependence on stellar masses. This can reduce the luminosities and photon fluxes 
of helium by factors $\sim 4-8$, depending on the sensibility to the stochastic effects 
of the IMF considered.

 For simplicity, we have used in our photo-ionization modelling a constant nebular 
hydrogen density. We note, however, that since we extend our calculations to large 
time scales, these density values might very well change; e.g., supernova feedback 
from the most massive stars, stellar winds, changes in the ionizing radiation field, etc. 
As mentioned before, high nebular densities favor high line luminosities, specially for 
helium lines, so a decrease of density due to the above processes would be 
accompanied by a decrease of luminosities at late times. This might be further 
considered in more detail making use of numerical simulations and radiative transfer 
codes which can engulf all the necessary physical processes. 

  We have explored the cases with average star formation rates ${\rm SFR}= 10^{-3}$ and 
${\rm SFR}= 10^{-4}$ M$_\odot \, {\rm yr^{-1}}$. These values are in agreement with the 
simulations by, e.g., \cite{Wise2012,Wise2014}, but these and other recent studies also 
allow for lower SFR. \cite{Cen2016} argues that the episodic bursts of star formation in atomic 
cooling halos are of the order of one every $20 - 100$ Myr, yielding ${\rm SFR}\sim10^{-5}$
M$_\odot \, {\rm yr^{-1}}$ or less, depending on the target mass of the bursts. In these 
cases, the high luminosity and photon flux values that we have obtained would only be 
present for a few Myr, during the life of the most massive stars. 

  Pop III stars are usually thought to exist only at very high redshifts, 
$z>11-25$, but various studies indicate that they can also form at much lower redshifts, up to 
$z\sim6-7$ \citep{Scannapieco2003,Salvadori2007,Tornatore2007b,Trenti2009,Wise2012,
Maio2013,Muratov2013,Visbal2016}. Recent simulations by \cite{Xu2016} find a non-negligible 
number of galaxies containing Pop III stars at $z\sim7$. These stars reside in dark matter 
halos of masses $M_h\sim10^7-10^8{\rm M_{\odot}}$ and the total stellar mass covers the 
range from a few tens to $\sim 1000$ M$_\odot$, in agreement with our target masses. 
In some cases, the mean metallicity of the halo is high enough to affect our luminosity results, 
especially for helium emission lines, but it is also possible that 
the regions where the stars reside contain only pristine gas \citep{Whalen2008,Xu2016}. 
If the metal enrichment by previous stars affects the star forming region, this could also  
supress the formation of successive metal-free objects, thus reducing the SFR, or even 
allowing only for an initial burst \citep{Nomoto2006,Heger2002}.

  It is important to mention that the maximum photon flux and luminosities that  
we have found correspond to the upper tails of the stochastic distributions. When these 
distributions are broad, specially for m9M500a2, the expected number of stochastic stellar 
populations giving rise to these values is small and the values 
in the peaks of the stochastic distributions are far from those in the tails. For 
this IMF, and considering populations with ages $>10$ Myr in Figure \ref{fig:csfr1000_Myr_L}, 
we find that $\sim0.2\%$ ($\sim33\%$) of the galaxies have luminosities above 
$L_{\rm \alpha}= 6.5\times10^{40}$ ($L_{\rm \alpha}= 4.5\times10^{40}$) ${\rm erg\, s^{-1}}$, 
and  $\sim0.3\%$ ($\sim22\%$) have values above $L_{\rm 1640}= 10^{39}$ 
($L_{\rm 1640}= 5\times10^{38}$) ${\rm erg\, s^{-1}}$. 
The probability for obtaining these extreme values increases and 
the peaks of the distributions are closer to the tails (the distributions narrow) when considering  
flatter IMFs. The simple deterministic calculation for m9M500a0 considering 
Ly$\alpha$ case-B departures and a total stellar mass 1000 M$_\odot$  
(Equation \ref{eq:lyab}) gives luminosities $L_{\rm \alpha}\sim 5\times10^{40}\,{\rm erg\, 
s^{-1}}$  and $L_{\rm 1640}\sim 10^{39}\,{\rm erg\, s^{-1}}$, which are similar to those for   
m9M500a2. The differences between tail and peak values are less important for the other 
populations with lower upper mass limits.

  Many works have investigated the possible detection of Population III galaxies and 
the use of different techniques to distinguish them from the more `normal' galaxies 
\citep[see, e.g.,][and references therein]{Schaerer2014}.
This is a difficult task since different factors which are still not fully constrained (e.g., 
dust, galactic feedback, neutral gas fraction, gravitational 
lensing, etc.) can contribute to the enhancement or reduction of the 
intrinsic fluxes during the transmission through the ISM and IGM 
\citep[e.g.,][]{Johnson2009,Dijkstra2011,Laursen2011,Smith2015}. 
The resonant nature of the Ly$\alpha$ transition adds an extra difficulty in this case, 
strongly affecting the highly uncertain escape fraction of Ly$\alpha$ photons 
\citep[e.g.,][]{Dijkstra2014}. We consider the simple case of our default target mass, 
1000 M$_\odot$, with intrinsic luminosity values $L_{\rm \alpha}\sim 5\times10^{40}\,
{\rm erg\, s^{-1}}$  and $L_{\rm 1640}\sim 10^{39}\,{\rm erg\, s^{-1}}$. We calculate the 
fluxes expected from populations at redshift $z=7,\,10$ and $14$, accounting for the 
dilution factor due to the luminosity distance, and we compare 
them with the sensitivity limit of the Near Infrared Spectrograph (NIRSpec) aboard the 
JWST \citep{Gardner2006}. We consider the case of line fluxes from a point source, with 
a signal-to-noise, ${\rm SNR}=10$, and $10^4$ s of integration 
time\footnote{\url{http://www.stsci.edu/jwst/science/sensitivity/spec1.jpg}}. We find that 
Ly$\alpha$ luminosity is a factor of $\sim20$ below the detection threshold at redshifts 
$z=7$ and $10$, and falls below a factor of $40$ at $z=14$. The helium luminosity is a 
factor $\sim450$ below the detection limit at $z=7$ and almost three orders of magnitude 
below for the highest redshifts. Therefore, higher total stellar masses and/or 
gravitational lensing \citep{Stark2007,
Zackrisson2012,Zackrisson2015} appear to be indispensable for a spectroscopic 
confirmation of Pop III galaxies (non-metal detection and the presence of strong Ly$\alpha$ 
emission), and also for detection with, e.g., the combination of wide-field surveys and 
multiband photometry \citep[see, e.g.,][]{Zackrisson2011b,Zackrisson2011a,Inoue2011}. 


\section{Summary and conclusions}\label{sec:summary}

  We have revisited the spectra of population III galaxies, specifically focusing on 
the nebular hydrogen Ly$\alpha$ and ${\rm HeII}\,\lambda1640$ lines. 
We have computed a series of stellar helium and hydrogen model atmospheres covering 
a mass range from ${\rm 9\,to\,1000\,M_{\odot}}$.
We have used the stars to construct several populations, allowing for 
top-heavy and Salpeter slope IMFs, and different upper stellar mass limits. 
We have included upper mass limits that are lower than those in earlier works, motivated by 
recent studies favoring the formation of low mass objects, and, for the first time in studies 
of Pop III populations, we have considered the stochastic sampling of the IMF.  
We have obtained the nebular spectra using our own SEDs and the 
photoionization code \textit{Cloudy}. Finally, we have explored the departures from the 
case-B recombination assumption for our populations, and we have revisited their origin. 
Our results can be summarized as follows.

\begin{itemize}[leftmargin=0pt,itemindent=20pt]
\item The Ly$\alpha$ line flux is enhanced by a factor of $2 - 3$ compared to case-B, depending 
on the IMF and parameters of the nebula, in agreement with \cite{Raiter2010}. 
Our analysis shows that the origin of the Ly$\alpha$ departures from 
case-B is due mainly to energetic free electrons which collisionally excite and ionize 
hydrogen atoms. The photoionization from the first excited state, argued to be the major 
mechanism in previous works, is negligible.

\item Stochastic sampling of the IMF produces large fluctuations in the ionizing photon 
flux, specially for the single-ionized helium (Q(HeII); more than a factor 100 for a total stellar 
mass of 1\,000 M$_\odot$), due to the strong dependence 
on the mass of the stars within the populations. For the case of hydrogen and the LW band, the 
distributions around the deterministic value are much narrower. The stochastic effects are 
more important for IMFs with high upper mass limits and also for those which disfavor 
the presence of massive stars, i.e., those with Salpeter slopes instead of flat IMFs. 
Stochasticity can enhance the Ly$\alpha$ flux up to a factor 
$\sim 3$ in populations with massive stars (upper mass limit IMF of 500 M$_\odot$). This, 
added to the case-B departure implies a boost by a factor of $\sim 8$ compared 
to the common calculations. For populations with less massive stars the total boost reaches 
values $\sim$5, depending on the IMFs and nebular parameters. For the case of 
the ${\rm HeII}\,\lambda1640$ line, stochasticity boosts the flux up to a factor $\sim 3$.
Accounting for a lower total stellar mass, 100 M$_\odot$, the stochastic effects increase 
significantly showing the strong anti-correlation between these two parameters. In this case, 
the fluctuations found for the hydrogen (LW) ionizing photon fluxes can have implications 
to the ionizing (dissociating) power inferred for Pop III stars. On the other 
hand, for a total stellar mass of 10\,000 M$_\odot$, the stochastic effects 
are strongly reduced.

\item{When considering the effect of time evolution for a single starburst we see that the 
distributions slightly broaden with time, specially for the case of flat IMFs. The stochastic 
effects are again more important for high stellar mass upper limit IMFs and for low values of the 
total stellar mass. Considering periodic bursts of star formation, we observe similar  
stochastic behaviours as before, and all the distributions flatten after $\lsim 10$ Myr. 
The distributions reach maximum intrinsic luminosity values of $L_{\rm \alpha}\sim 
5\times10^{40}\,{\rm erg\, s^{-1}}$  and $L_{\rm 1640}\sim 10^{39}\,{\rm erg\, s^{-1}}$ 
for a total stellar mass of 1\,000 M$_\odot$. For the case of Ly$\alpha$, these values are a factor 
$\sim20$ below the sensitivity limits of the JWST at redshifts $z=7-10$, but this can improve 
considering the possible effects of gravitational lensing and/or higher total stellar mass. 
\cite{Visbal2016} have recently presented an analysis which 
allows for the formation of late ($z\sim7$), massive ($>10^6$ M$_{\odot}$) Pop III 
starbursts in $\sim 10^9$ M$_{\odot}$ dark matter halos,  
due to photoionization feedback effects by nearby galaxies. In addition, 
these stellar populations would likely reside in large ionized bubbles, thus 
illustrating that the neutral IGM may occasionally have a minor impact on the 
Ly$\alpha$ flux from Pop III galaxies \cite[see][]{Stark2016}.
Stellar populations of such characteristics would be above the detection limits, 
even for the case of helium fluxes.}
\end{itemize}

We conclude by stressing that the joint effect of the stochastic sampling of the IMF and 
case-B departures analysed in this work can give rise to values of the Ly$\alpha$ and 
HeII$\, \lambda1640$ line fluxes which deviate significantly from the standard analytical 
calculations which ignore these effects. Our study, as well as previous works, shows   
that gravitational lensing is required to detect Pop III galaxies with JWST. However, the 
enhancement obtained in our results reduces significantly the required magnification.

\acknowledgements
We thank the anonymous referee for constructive comments that 
helped to improve significantly the discussion about our model 
parameters and results.
We thank Shinya Miyake, Mikael Toresen, Mats Carlsson, Max Gronke 
and Marit Sandstad for their ideas and comments in the early 
stages of this project. We also thank, for their useful opinions, the people in 
the conference `First stars, galaxies and black holes; now and then' held  
in Groningen in June 2015. Specially Anna Schauer, Eli Visbal, Athena Stacy, 
Kevin Schlaufman and Ylva Gotberg, among others. We are grateful to Peter 
Schilke and Kohei Inayoshi for sharing their thoughts about the radiative processes 
governing the HII regions. Finally, we thank Daniel Stark, Masami Ouchi, Andrea Ferrara, 
Avery Meiksin, Nick Gnedin and Jeff Cooke for pointing several important observational 
aspects. MD thanks the astronomy department at UC Santa Barbara for their kind hospitality.


\bibliographystyle{apj}
\bibliography{stellar}\label{References}

\end{document}